\documentclass[useAMS,usenatbib]{mn2e}

\usepackage{amssymb,amsmath}
\usepackage{graphicx}
\usepackage{natbib}
\usepackage{aas_macros}

\newcommand{\hii}{H~{\sc ii}}

\begin{document}

\title[The SILCC project IV]{The SILCC project --- IV. Impact of dissociating and ionising radiation on the interstellar medium
and H$\alpha$ emission as a tracer of the star formation rate}

\author[Thomas Peters et al.]{
\parbox[h]{\textwidth}{
Thomas Peters$^1$\thanks{E-Mail: tpeters@mpa-garching.mpg.de},
Thorsten Naab$^1$,
Stefanie Walch$^2$,
Simon C. O. Glover$^3$,
Philipp Girichidis$^1$,
Eric Pellegrini$^3$,
Ralf S. Klessen$^{3,4}$,
Richard W\"{u}nsch$^5$,
Andrea Gatto$^1$,
Christian Baczynski$^3$
}\vspace{0.4cm}\\
\parbox{\textwidth}{$^1$Max-Planck-Institut f\"{u}r Astrophysik, Karl-Schwarzschild-Str. 1, D-85748 Garching, Germany\\
$^2$I. Physikalisches Institut, Universit\"{a}t zu K\"{o}ln, Z\"{u}lpicher Strasse 77, D-50937 K\"{o}ln, Germany\\
$^3$Universit\"{a}t Heidelberg, Zentrum f\"{u}r Astronomie, Institut f\"{u}r Theoretische Astrophysik, Albert-Ueberle-Str. 2, D-69120 Heidelberg, Germany\\
$^4$Universit\"at Heidelberg, Interdisziplin\"{a}res Zentrum f\"{u}r Wissenschaftliches Rechnen (IWR), D-69120 Heidelberg, Germany\\
$^5$Astronomical Institute, Academy of Sciences of the Czech Republic, Bo\v{c}n\'\i\ II 1401, 141 31 Prague, Czech Republic
}}

\maketitle

\begin{abstract}
We present three-dimensional radiation-hydrodynamical simulations of the impact of
stellar winds, photoelectric heating, photodissociating and photoionising radiation, and
supernovae on the chemical composition and star formation in a stratified disc
model. This is followed with a sink-based model for star clusters with populations
of individual massive stars. Stellar winds and ionising radiation regulate the star
formation rate at a factor of $\sim10$ below the simulation with only supernova feedback
due to their immediate impact on the ambient interstellar medium after star
formation. Ionising radiation (with winds and supernovae) significantly reduces the
ambient densities for most supernova explosions to $\rho < 10^{-25}\,$g\,cm$^{-3}$,
compared to $10^{-23}$g\,cm$^{-3}$ for the model with only
winds and supernovae. Radiation from massive stars reduces the amount of molecular
hydrogen and increases the neutral hydrogen mass and volume filling fraction. Only
this model results in a molecular gas depletion time scale of $2\,$Gyr and shows the
best agreement with observations.
In the radiative models, the H$\alpha$ emission is dominated by radiative
recombination as opposed to collisional excitation (the dominant emission in
non-radiative models), which only contributes $\sim1$--$10\,$\% to the total H$\alpha$
emission.
Individual massive stars ($M \geq30\,$M$_\odot$) with short
lifetimes are responsible for significant fluctuations in the H$\alpha$ luminosities.
The corresponding inferred star formation rates can underestimate the true
instantaneous star formation rate by factors of $\sim10$. 
\end{abstract}

\section{Introduction}
\label{sec:intro}

Massive stars (O and B stars with masses in excess of $8\,$M$_{\odot}$)
dominate the energy output of newly formed stellar populations. Most
of the energy is emitted in the form of (ionising) radiation, followed
by supernova explosions (about one order of magnitude less) and
stellar winds (another order of magnitude less).
Photoionising radiation (photon energies larger than $13.6\,$eV) is a major source of ionised hydrogen and
drives the formation of \hii\ regions
(e.g. \citealp{1979MNRAS.186...59W,2005MNRAS.358..291D,2010ApJ...711.1017P,2012MNRAS.427..625W,2015MNRAS.448.3248G}),
which cool by Lyman $\alpha$ radiation. This effect is significant 
although, energetically, the coupling of gas and radiation is usually
very inefficient (less than $0.1\,$\% of the emitted energy in the Lyman continuum
can be converted into kinetic energy, see
e.g. \citealp{2012MNRAS.427..625W}). Photoionising radiation is also a
major source for H$\alpha$ emission, which is used as one of the major
tracers for star formation in galaxies at low and high redshifts
(e.g. \citealp{kenn98,2009ApJ...706.1364F}) 

How radiation couples to the surrounding gas depends on the wavelength
of the radiation, and energies below the Lyman continuum also have to be
considered. For example, photoelectric heating of dust 
(photon energies of $5.6\,$eV -- $13.6\,$eV) is the dominant heat
source in the interstellar medium \citep[ISM,][]{draine78}. It can lead to temperatures of
a few $10^3\,$K and contribute to the formation of the warm, neutral
component of the ISM (WNM), where the fine-structure 
lines [C~II] and [O~I] are the main coolants. In addition,
photodissociating radiation (photon energies of $11.2\,$eV -- $13.6\,$eV)
will destroy molecular hydrogen and change the abundance
ratios. Photoionising radiation from massive stars can heat the ISM to
temperatures up to $\sim10^4\,$K and impacts the chemistry and 
thermodynamics of the cold neutral, the warm neutral, and the warm
ionised medium directly. The cold neutral medium (in
particular molecular hydrogen) is the fuel for new stars and
correlates well with the star formation rate (SFR) of galaxies at low
as well as high redshifts (e.g. \citealp{bigiel08,2010Natur.463..781T}). 

Supernovae and to some degree stellar winds are energetic enough to
shock-heat the ISM to temperatures of $\sim10^6\,$K  and generate hot gas
\citep{1977ApJ...218..377W,1988RvMP...60....1O} for which
bremsstrahlung emission becomes the dominant cooling radiation. It has
been argued that supernova-driven shocks play a significant role for
driving ISM turbulence \citep{2004RvMP...76..125M}, accumulate dense and cold 
gas and form a hot, possibly volume-filling, medium which is
responsible for driving galactic outflows (see e.g. \citealt{2006ApJ...653.1266J}).   

The emission of ionising radiation, stellar winds and supernova
explosions
are therefore the dominant sources 
that determine
the chemo-thermodynamic properties of the ISM. They should
all be considered in modern attempts to improve the numerical
modelling towards a more complete model of the ISM.   

Significant progress has been made on individual aspects. Simulations
focusing on the impact of supernovae on ISM structure
\citep{2015ApJ...802...99K,2015MNRAS.451.2757W,2015MNRAS.450..504M}
indicate a regulation of star formation and vertical disc structure in models
limited to momentum injection (e.g. \citealp{2015ApJ...815...67K}) as
well as the formation of galactic outflows in simulations including
the formation of a hot phase
(e.g. \citealp{1999ApJ...514L..99K,2006ApJ...653.1266J,2004A&A...425..899D,2015ApJ...800..102H,2015ApJ...813L..27P,girietal16}).  
Stellar winds impact the ambient ISM structure, possibly terminating
gas accretion onto star-forming regions and thereby regulating the
local efficiency of star formation \citep{gatto16}. They also change the conditions
for subsequent supernova explosions by reducing the ambient gas
densities making energy deposition from supernovae more efficient.
Ionising radiation has a similar effect by heating the dense gas phase
in star-forming regions to $10^4\,$K, injecting momentum into the ISM,
and driving turbulence locally \citep{petersetal08,2009ApJ...694L..26G,2015MNRAS.448.3248G}.  

In this paper we present the first chemo-dynamical numerical models
sequentially including all the above processes---ionising radiation
followed with the radiation transfer code \texttt{Fervent} \citep{baczynski15},
stellar winds, and supernova explosions---in combination with a sink
particle-based star cluster formation model \citep{gatto16}. We investigate a small
region of a galactic disc with solar neighbourhood-like properties in
the stratified disc approximation. The model is combined with a
chemical network to follow the evolution of molecular, neutral and
ionised gas in the presence of an external radiation field (see
\citealp{walchetal15} for details on the SILCC project,
https://hera.ph1.uni-koeln.de/$\sim$silcc/). In this paper we do not
investigate the effect of non-thermal ISM constituents like 
magnetic fields or cosmic rays.  

The paper is organised as follows. In Section~\ref{sec:sim} we present
an overview of the simulations followed by a discussion about the
energy budget of wind, radiation and supernova feedback processes
(Section \ref{sec:bud}). A qualitative discussion of the simulation
results (Section \ref{sec:qual}) is followed by quantitative analyses of the
star cluster properties (Section \ref{sec:stcl}), energy input
(Section~\ref{sec:enin}), mass fraction in different ISM phases
(Section~\ref{sec:mf}), depletion times (Section~\ref{sec:dep}) and volume filling
fractions (Section~\ref{sec:vff}). The origin of H$\alpha$ emission is discussed
and interpreted in 
in Sections~\ref{sec:halpha}, \ref{sec:lsd}, and \ref{sec:cal}. We
conclude in Section~\ref{sec:con}.

\section{Simulations}
\label{sec:sim}

We use the adaptive mesh refinement code \texttt{FLASH}~4 \citep{fryxell00,dubey09}
to run kpc-scale stratified box simulations. We employ a stable, positivity-preserving
magnetohydrodynamics solver \citep{bouchut07,waagan09,waagan11}. Self-gravity is
incorporated with a Barnes-Hut tree method (R.~W\"{u}nsch et al. in prep.).
The simulated boxes have dimensions $0.5\,$kpc $\times 0.5\,$kpc $\times 10\,$kpc.
The box boundaries are periodic in the disc plane ($x$ and $y$ directions).
We allow the gas to leave the simulation box in the vertical ($z$) direction, but prevent
infall into the box from the outside (diode boundary conditions). The highest grid
resolution is $\Delta x = 3.9\,$pc.

The gravitational force of the stellar component of the gas is modelled with an
external potential. We assume an isothermal sheet potential with a stellar surface
density $\Sigma_{*} = 30\,$M$_\odot\,$pc$^{-2}$ and a vertical scale height $z_\mathrm{d} = 100\,$pc.
These parameters were chosen to fit solar neighbourhood values. The gas is set up
with a surface density $\Sigma_\mathrm{gas} = 10\,$M$_\odot\,$pc$^{-2}$. In $z$-direction,
the gas follows a Gaussian distribution with a scale height of $60\,$pc.
We do not include magnetic fields in the simulations presented in this paper.
More information on the initial conditions and the setup of the simulations can be found in
\citet{walchetal15} and \citet{girietal16}.

Heating and cooling processes, as well as molecule formation and destruction, are treated
with a time-dependent chemical network \citep{nellan97,glovmcl07,glovmcl07b,glovetal10,glovclar12}.
The network follows the abundances of free electrons, H$^{+}$, H, H$_{2}$, C$^{+}$, O and CO.
Warm and cold gas primarily cools via Lyman-$\alpha$ cooling, H$_{2}$ ro-vibrational line cooling, fine-structure emission from C$^{+}$ and O, and
rotational emission from CO \citep{glovetal10,glovclar12}.
In the hot gas, we also take the electronic excitation of helium and of partially  
ionised metals into account following the \cite{gnafer12} cooling rates.
We assume X-ray ionisation and heating rates based on \citet{woletal95} and
a cosmic ray ionisation rate $\zeta = 3 \times 10^{-17}\,$s$^{-1}$ following \citet{gollan78}.
Furthermore, we assume a diffuse interstellar radiation field $G_0 = 1.7$ \citep{habing68,draine78}
and include the effect of heating from the photoelectric effect using the prescription by \citet{baktie94}.
Shielding by dust as well as molecular self-shielding is modelled with the 
\texttt{TreeCol} method (\citealt{clarketal12}, R.~W\"{u}nsch et al. in prep.).
The assumed metallicity is solar.
For more information on the chemical network and our description of heating and cooling see \citet{gattoetal15} and \citet{walchetal15}.

Star clusters form dynamically and are modelled using sink particles \citep{federrathetal10}.
The accretion radius of the sink particles is $r_\mathrm{acc} = 4.5 \times \Delta x = 17.58\,$pc.
The sink particles have a threshold density of $\rho_\mathrm{thr} = 2 \times 10^{-20}\,$g\,cm$^{-3}$.
All gas within $r_\mathrm{acc}$ that is above $\rho_\mathrm{thr}$, bound to the sink particle and
collapsing towards the centre will be removed from the grid and accreted onto the sink particle.
As soon as we have accreted $120\,$M$_\odot$ of gas onto a sink particle, we randomly sample a
massive star with a mass between $9$ and $120\,$M$_\odot$ from a Salpeter initial mass function \citep[IMF,][]{salpeter55}.
We follow the stellar evolution of each of these massive stars according to the \citet{ekstroem12}
tracks until they explode as supernovae.
We refer to \citet{gatto16} for a detailed description of our cluster sink particles.

The simulations contain three different types of feedback from the formed star clusters.
First, we include the radiative feedback from the massive stars in our stellar cluster sink particles.
The stellar evolutionary tracks provide luminosities and effective temperatures as a function of
stellar age, from which we compute the emitted spectrum.
We use the \texttt{Fervent} code \citep{baczynski15} to propagate the radiation
from all sources individually
in four different energy bins across the adaptive mesh using raytracing. 
The first energy bin $5.6\,$eV$< E_{5.6} < 11.2\,$eV is responsible for the photoelectric heating
of the ISM. Photons in the second energy bin $11.2\,$eV$< E_{11.2} < 13.6\,$eV
are energetic enough to photodissociate molecular hydrogen through the process
$\rm{H}_2 + \gamma_{11.2} \to \rm{H} + \rm{H}$. Photons from the third energy bin,
$13.6\,$eV$< E_{13.6} < 15.2\,$eV, can photoionise atomic hydrogen via 
$\rm{H} + \gamma_{13.6} \to \rm{H}^++ \rm{e}^-$. Photons in the fourth bin,
$15.2\,$eV$< E_{15.2}$, can still photoionise atomic hydrogen, but they are also able to photoionise
molecular hydrogen, $\rm{H}_2 + \gamma_{15.2^+} \to \rm{H}_2^++ \rm{e}^-$. Which one of these two
processes occurs for a photon in this bin is decided
based on the respective absorption cross sections if both forms of hydrogen are present in a given grid cell.
The $\rm{H}_2^+$ ions formed by the latter process are assumed to immediately undergo dissociative recombination,
resulting in the production of two hydrogen atoms. We do not include any form of radiation pressure.
For more details on the complex photochemistry included in the \texttt{Fervent} code we refer to the
method paper \citep{baczynski15}.

The second feedback process we include is the mechanical feedback from stellar winds.
We take the mass-loss rates $\dot{M}_\mathrm{wind}$ directly from the \citet{ekstroem12}
stellar evolutionary tracks. The computation of the wind terminal velocity $v_\mathrm{wind}$
depends on the evolutionary status of the stars. For OB stars and A supergiants,
we use the scaling relations from \citet{kudpul00} and \citet{marpul08} on both sides
of the bi-stability jump, and linearly interpolate in between \citep{pulsetal08}.
For WR stars, we linearly interpolate the observational data compiled by \citet{crowther07},
and for red supergiants we follow the scaling relation in \citet{vanloon06}.

Within a given cluster, we add up the contributions of all cluster members. The total wind luminosity
of a cluster with $N_\star$ stars is then
\begin{equation}
L_\mathrm{tot} = \frac{1}{2} \sum_{i = 1}^{N_\star} \dot{M}_{\mathrm{wind},i} \times v_{\mathrm{wind},i}^2 ,
\end{equation}
and the total mass-loss rate of the stellar cluster is
\begin{equation}
\dot{M}_\mathrm{tot} = \sum_{i = 1}^{N_\star} \dot{M}_{\mathrm{wind},i} .
\end{equation}
For each time step $\Delta t$, we add the mass $\dot{M}_\mathrm{tot} \times \Delta t$ to all cells within the
wind injection region. We set the radius of this region equal to the sink accretion radius $r_\mathrm{acc}$
and distribute the injected mass equally among all cells within $r_\mathrm{acc}$ from the location of the sink centre.
We assume that the wind is spherically symmetric and set the radial wind velocity of all cells within $r_\mathrm{acc}$
to
\begin{equation}
v_\mathrm{rad} = \sqrt{\frac{2 \times L_\mathrm{tot} \times \Delta t}{M_\mathrm{inj}}} ,
\end{equation}
where $M_\mathrm{inj}$ is the mass within the injection region. For more details on our implementation
of stellar wind feedback, we refer to \citet{gatto16}.

The third feedback process is the thermal feedback from supernova explosions. At the end of each massive
star's life, we inject a thermal energy $E_\mathrm{SN} = 10^{51}\,$erg into a spherical region of
radius $r_\mathrm{acc}$ around the sink particle. Additionally, we distribute the mass of the supernova
progenitor equally over the cells in this volume.
A detailed description of the supernova injection subgrid model is presented in \citet{gatto16}.

Initially, we create a complex density structure by driving turbulence with an external forcing.
This is necessary because otherwise the homogeneous disc would undergo global collapse and create a starburst.
We inject kinetic
energy with a flat power spectrum on the two largest modes in the plane of the disc, corresponding to the box
size and half of the box size. We apply a natural mixture of 2:1 between solenoidal (divergence-free) and
dilatational (curl-free) modes. The forcing field evolves according to an Ornstein-Uhlenbeck process \citep{eswapope88}
with an autocorrelation time of $49\,$Myr, which corresponds to the crossing time in $x$ and $y$ directions. The amplitude
of the forcing is adjusted such that a global, mass-weighted root-mean-square velocity of $10\,$km\,s$^{-1}$
is attained. We switch off the forcing with the formation of the first sink particle, which happens at $t = 30\,$Myr.

We use a notation for our simulations that is consistent with \citet{gatto16}. Run FSN only includes feedback
from supernova explosions, run FWSN incorporates feedback from winds and supernovae,
and run FRWSN integrates feedback from radiation,
winds and supernovae simultaneously. The simulations presented in this paper are summarised in Table~\ref{table:sim}.
We have run all simulations for a total time $t_\mathrm{max} = 70\,$Myr.

\begin{table}
\caption{Simulations}
\label{table:sim}
\begin{center}
\begin{tabular}{lccc}
\hline
simulation      & supernova & wind      & radiation \\
name            & feedback  & feedback  & feedback  \\
\hline
FSN             & yes       & no        & no  \\
FWSN            & yes       & yes        & no  \\
FRWSN           & yes       & yes       & yes  \\
\hline
\end{tabular}
\medskip\\
Overview of the simulations. We list the simulation names and the included feedback processes.
\end{center}
\end{table}

\section{Energy budget of the feedback processes}
\label{sec:bud}

\begin{figure}
\centerline{\includegraphics[width=\linewidth]{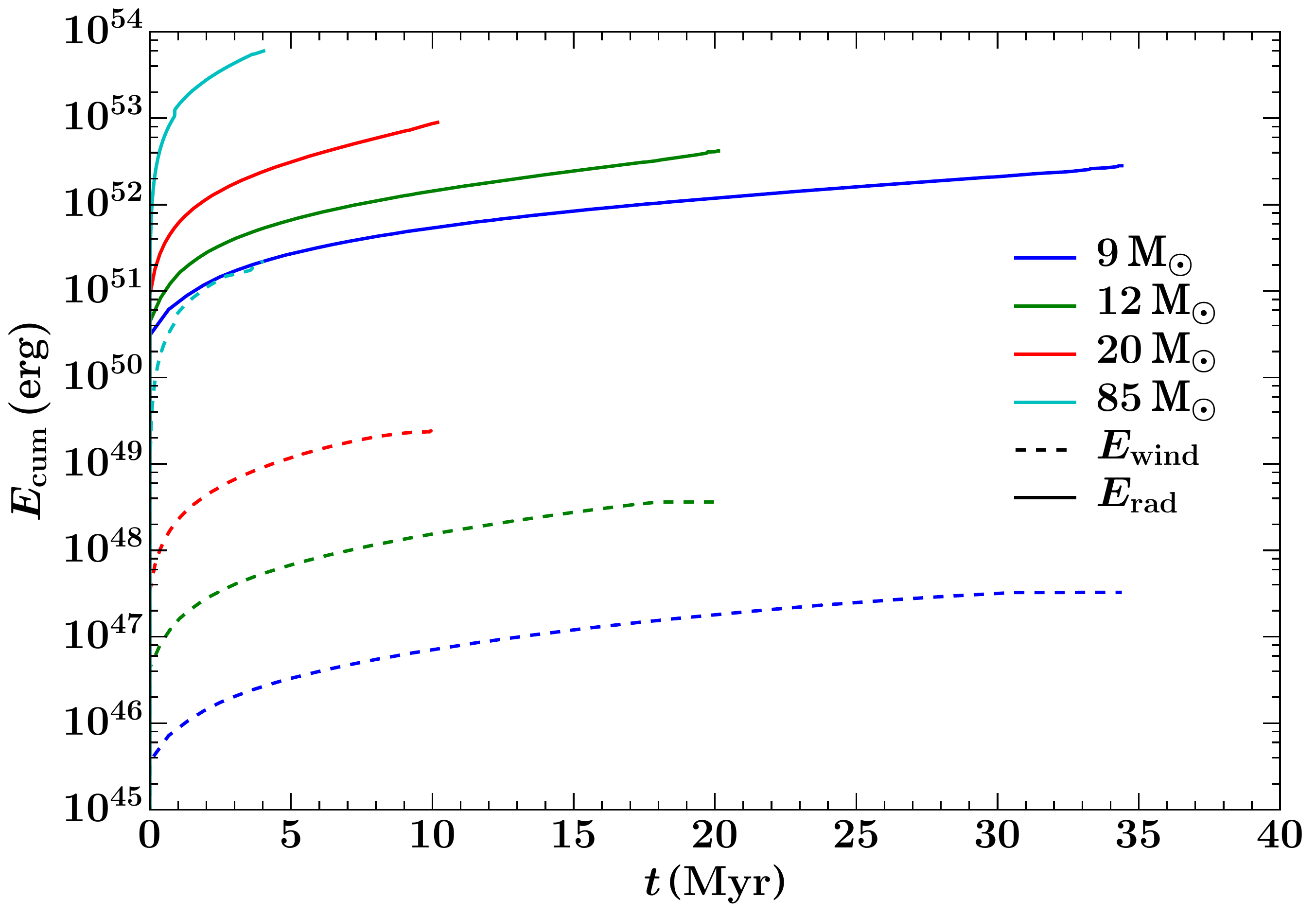}}
\centerline{\includegraphics[width=\linewidth]{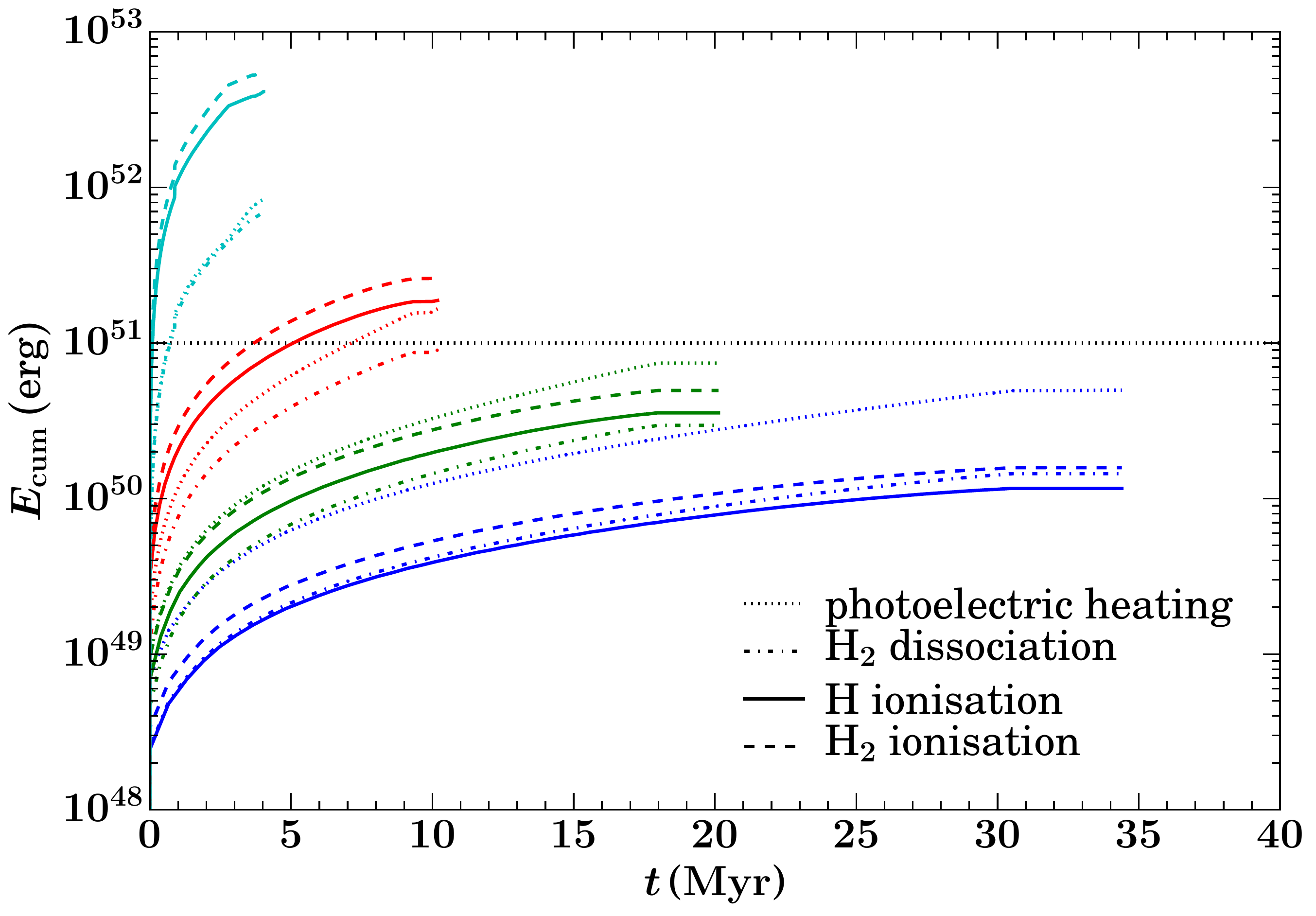}}
\caption{Cumulative energy released by different forms of feedback. \emph{(top)} The total radiation energy $E_\mathrm{rad}$
(solid lines) is $2$ to $5$ orders of magnitude larger than the wind energy $E_\mathrm{wind}$ (dotted lines), dependent on the stellar mass.
\emph{(bottom)} The radiation energy that can potentially couple to the gas is much smaller, but can be comparable
to the energy injected by a supernova explosion. The dominant photochemical processes are a function of stellar mass.
The horizontal black dashed line indicates the supernova injection energy $E_\mathrm{SN} = 10^{51}\,$erg.}
\label{fig:tracks}
\end{figure}

Before we start a differential analysis of the simulations, it is instructive to consider the energies associated
with the different forms of feedback. In Figure~\ref{fig:tracks}, we show the cumulative energy released by radiative and wind
feedback for a star with $9$, $12$, $20$ and $85\,$M$_\odot$. The stellar evolutionary tracks are identical to those shown
in Figure~1 of \citet{gatto16}.

The total radiation energy $E_\mathrm{rad}$ is computed from the stellar bolometric luminosities.
It is $2$ to $5$ orders of magnitude larger than the wind energy $E_\mathrm{wind}$, dependent on the stellar mass.
The higher the stellar mass, the smaller is the difference in the energies. But in contrast to $E_\mathrm{wind}$,
$E_\mathrm{rad}$ is not fully deposited into the gas around the source. Photons with energies below $5.6\,$eV do not
couple to the gas at all. More energetic photons only transfer energy to the gas in the presence of a sufficiently high column
of absorbers. And even in this case, a significant fraction of energy is lost in overcoming the binding energies
of the photochemical processes. Figure~\ref{fig:tracks} also shows the maximum amount of energy that can be deposited
into the gas within the \texttt{Fervent} energy bands, taking the binding energies into account and assuming high optical
depths in all directions.
For the $5.6\,$eV$< E_{5.6} < 11.2\,$eV energy band, we assume a photoelectric heating efficiency of $5\,$\%, which
is the largest value we expect to encounter in the dense ISM \citep{baktie94}.\footnote{In practice, the effective
efficiency may be a factor of a few smaller than this. For example, in M31 it is around $2\,$\% (Kapala et al., in prep.).}
In this case, the cumulative energies are orders of magnitude smaller than $E_\mathrm{rad}$,
but still significantly larger than $E_\mathrm{wind}$. For the $9$ and $12\,$M$_\odot$ stars, the majority of
radiation energy is released by photoelectric heating, while for the $20\,$ and $85\,$M$_\odot$ stars,
photoionisation of H$_2$ and H are dominant.
The cumulative energy available to photoelectric heating of a $12\,$M$_\odot$ star is comparable to the thermal energy
injected with a supernova explosion. For a star as massive as $20\,$M$_\odot$, the energy available in each of the four
energy bands is equivalent to the supernova explosion energy. If this radiation energy can be effectively
absorbed by the surrounding medium, we can expect a significant impact of the radiative feedback on the ISM.

\section{Qualitative discussion}
\label{sec:qual}

\begin{figure*}
\centerline{\includegraphics[width=\linewidth]{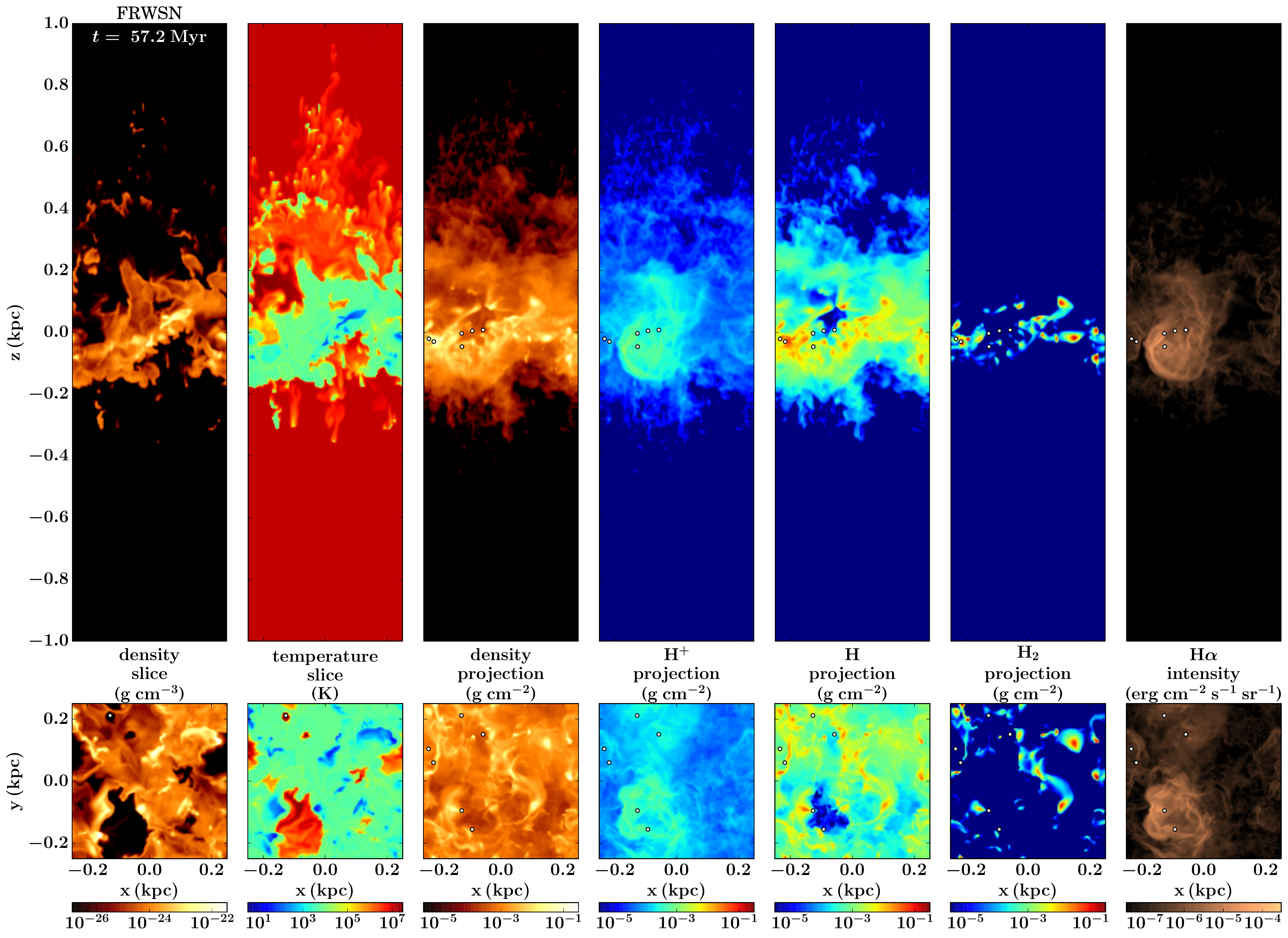}}
\caption{
Example snapshot from run FRWSN including radiation, stellar winds and supernovae at $t = 57.2\,$Myr. 
The picture shows edge-on (top) and face-on (bottom) density and temperature slices through the centre of the simulation box, projections of total
gas density and of the different forms of hydrogen (H$^+$, H and H$_2$), and an image of the resulting H$\alpha$ emission (from
left to right). The locations of the star
cluster particles are indicated with white circles. In the vertical direction, we only show the inner $2\,$kpc of the total $10\,$kpc
box height.
A movie of this simulation is available on the SILCC website (http://hera.ph1.uni-koeln.de/$\sim$silcc/).}
\label{fig:frwsn}
\end{figure*}

The initial $30\,$Myr of evolution are identical for all simulations. The runs only start to differ when star formation and stellar feedback
sets in.
Figure~\ref{fig:frwsn} shows an example snapshot\footnote{An animated version of this figure
can be found on the SILCC project website, http://hera.ph1.uni-koeln.de/$\sim$silcc/.}
from run FRWSN at $t = 57.2\,$Myr. This particular point in time was chosen because
it features a giant \hii\ region, allowing us to judge the impact of radiative feedback in comparison to the other simulations
without radiation. The picture shows edge-on (top) and face-on (bottom) density and temperature slices through the centre of the simulation box, projections of total
gas density and of the different forms of hydrogen (H$^+$, H and H$_2$), and an image of the resulting H$\alpha$ emission (from
left to right). The generation of the H$\alpha$ image is described in Section~\ref{sec:halpha}. The locations of the star
cluster particles are indicated with white circles. In the vertical direction, we only show the inner $2\,$kpc of the total $10\,$kpc
box height.

\begin{figure*}
\centerline{\includegraphics[width=\linewidth]{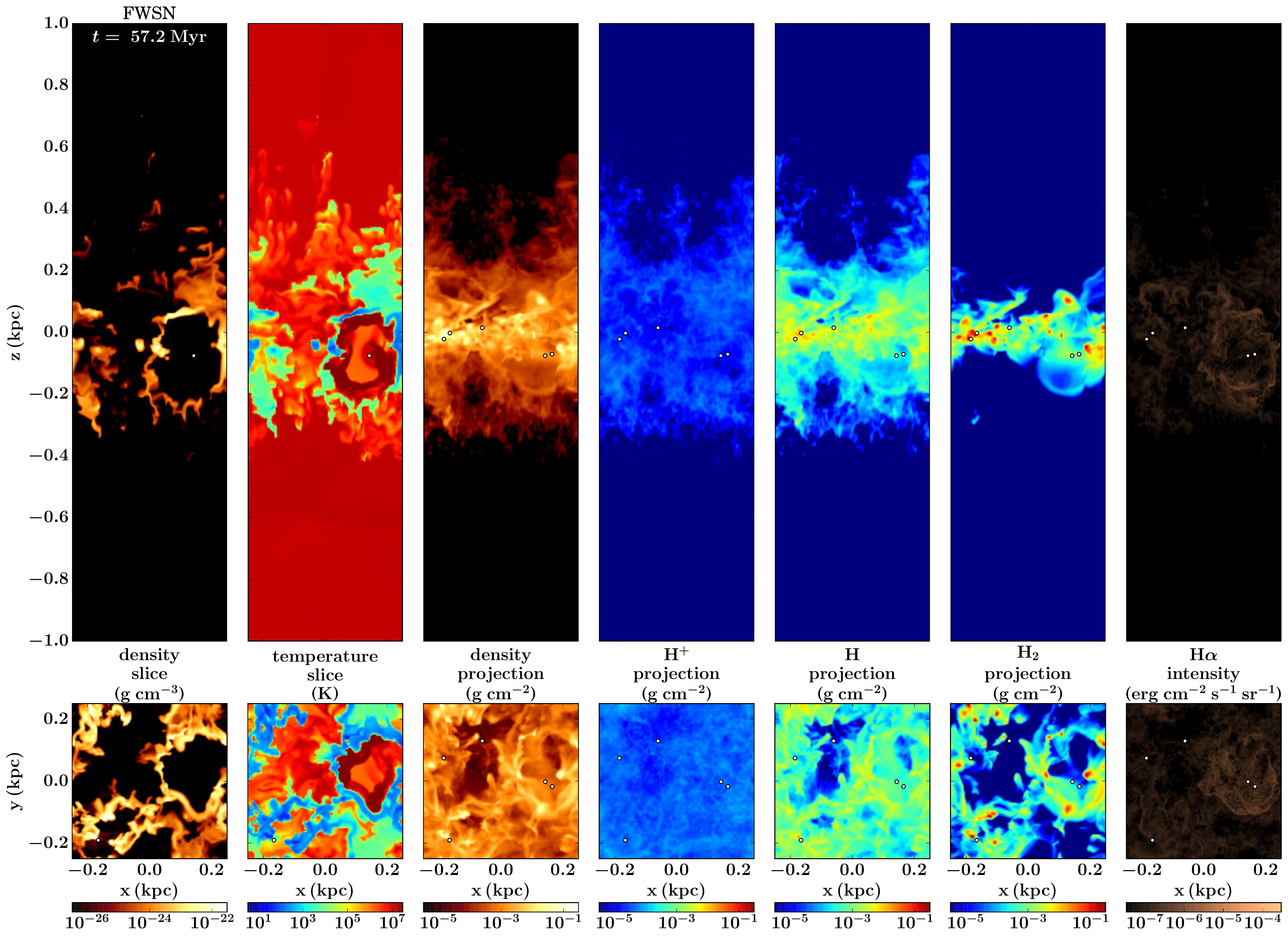}}
\caption{Same as Figure~\ref{fig:frwsn} for run FWSN including stellar winds and supernovae.
A movie of this simulation is available on the SILCC website (http://hera.ph1.uni-koeln.de/$\sim$silcc/).}
\label{fig:fwsn}
\end{figure*}

For comparison, Figure~\ref{fig:fwsn} shows run FWSN at the same time. It is clearly visible in the projections that the simulation box contains much
less H$^+$. On the other hand, the temperature slices reveal that the ionised gas in the disc is much hotter on average, $T > 10^5\,$K
instead of $T \approx 10^4\,$K for run FRWSN. As a result of both observations, the H$\alpha$ emission is much reduced compared to
run FRWSN (see Sections~\ref{sec:halpha} and~\ref{sec:lsd} for a detailed discussion). The number of sink particles in both simulations
is very similar. The total disc scale height is also comparable.

\begin{figure*}
\centerline{\includegraphics[width=\linewidth]{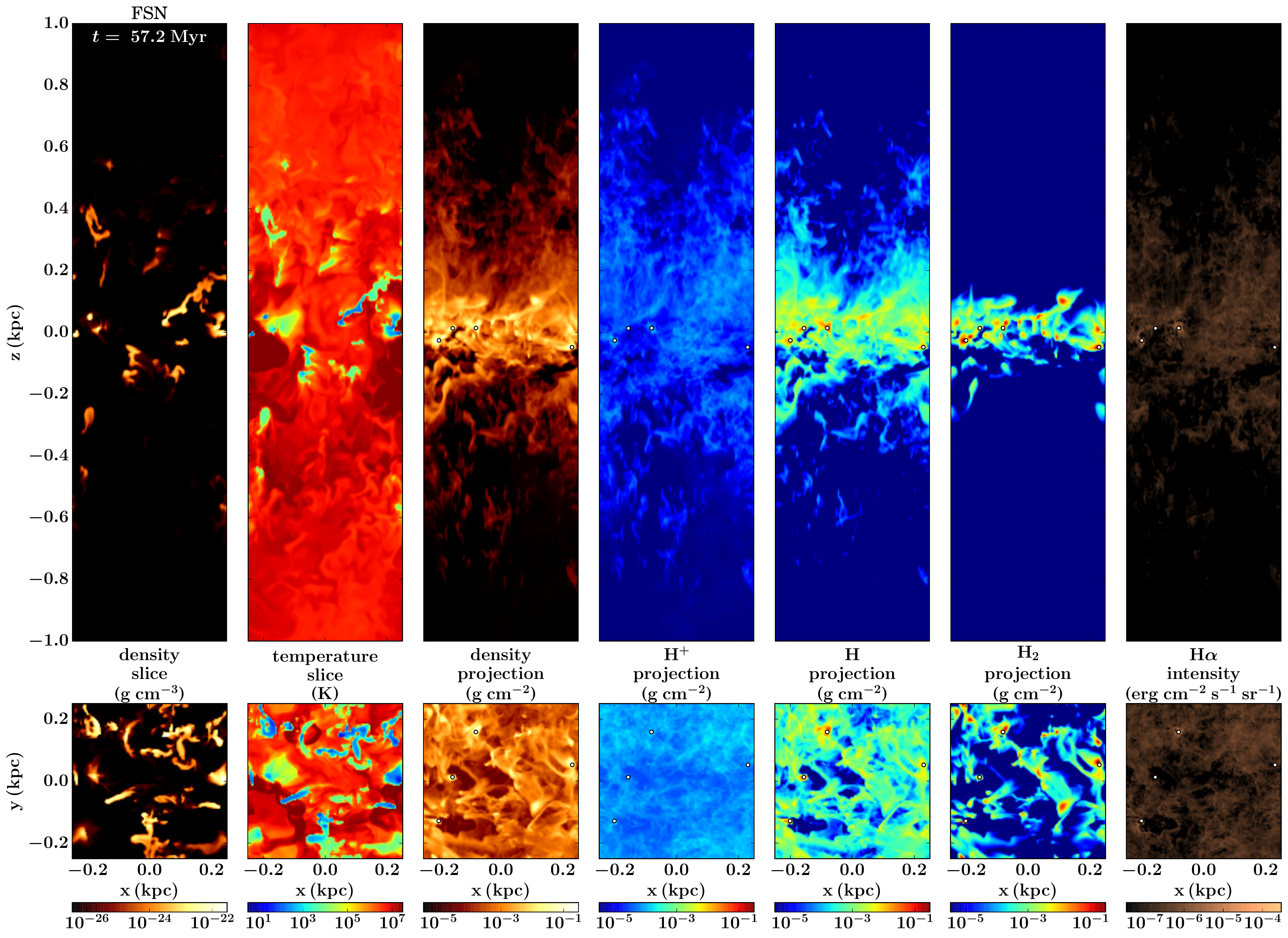}}
\caption{Same as Figure~\ref{fig:frwsn} for run FSN including only supernovae.
A movie of this simulation is available on the SILCC website (http://hera.ph1.uni-koeln.de/$\sim$silcc/).}
\label{fig:fsn}
\end{figure*}

In contrast, run FSN has a significantly larger disc scale height at the same time as the other two simulations, as we show in Figure~\ref{fig:fsn}.
There is even more hot gas present than in run FWSN. Because of the larger column of ionised gas, run FSN has a slightly enhanced H$\alpha$ emission
compared to run FWSN. Since the simulation FSN is the only run that produces a volume filling fraction of hot gas in excess of $50\,$\% (see Section~\ref{sec:vff}),
only this simulation drives a significant galactic outflow over an extended period of time \citep{gatto16}.

\section{Star clusters}
\label{sec:stcl}

\begin{figure}
\centerline{\includegraphics[width=\linewidth]{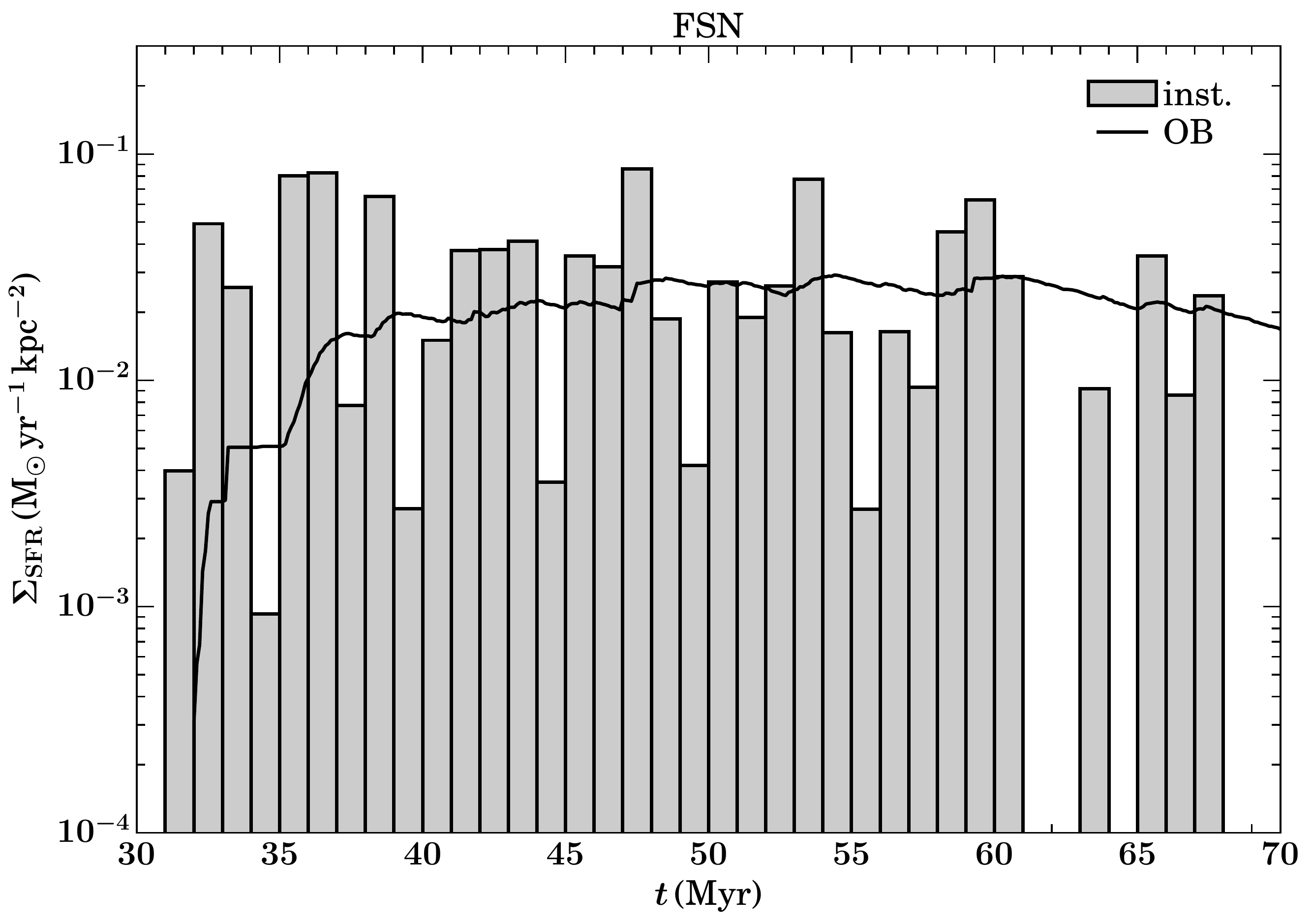}}
\centerline{\includegraphics[width=\linewidth]{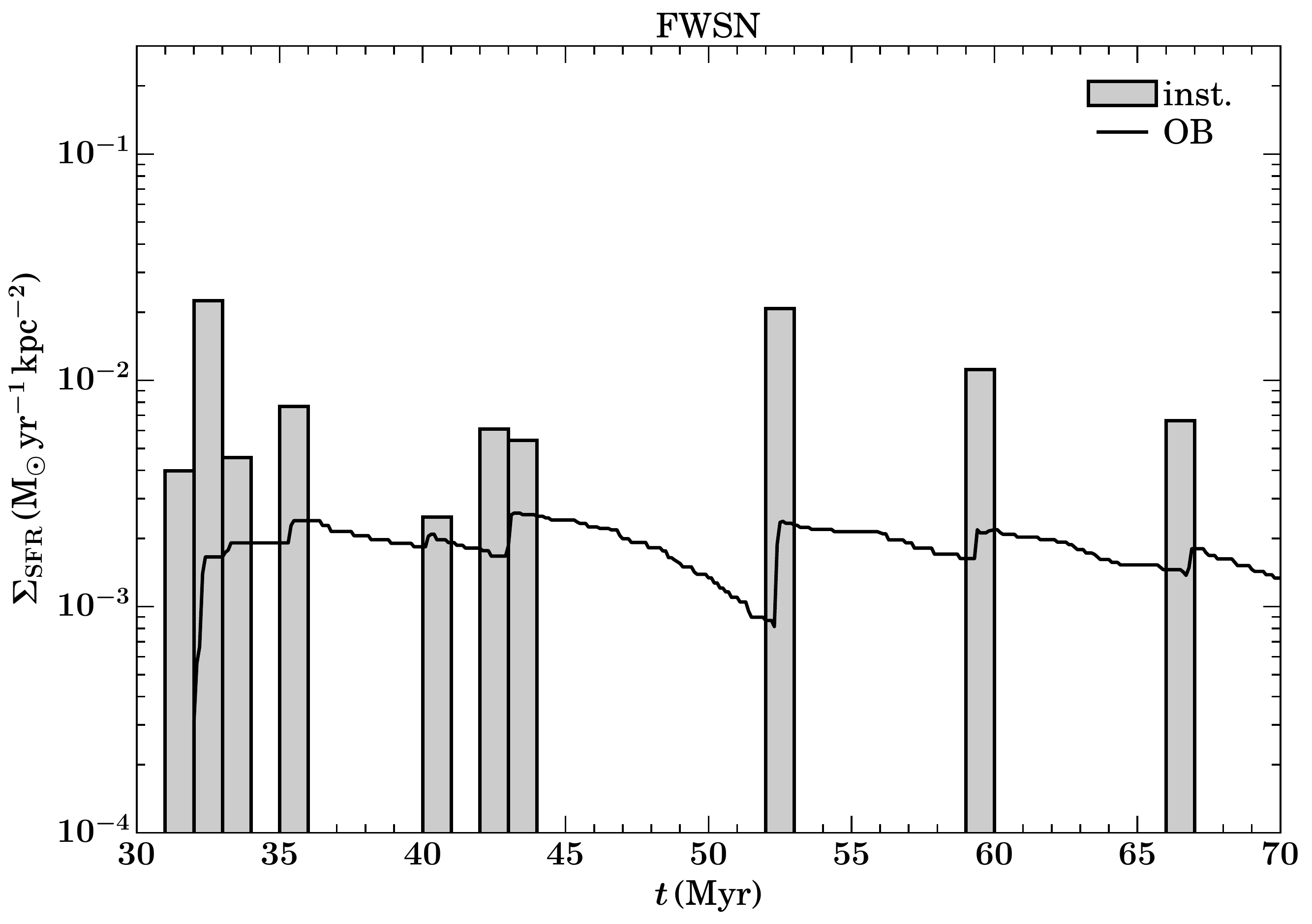}}
\centerline{\includegraphics[width=\linewidth]{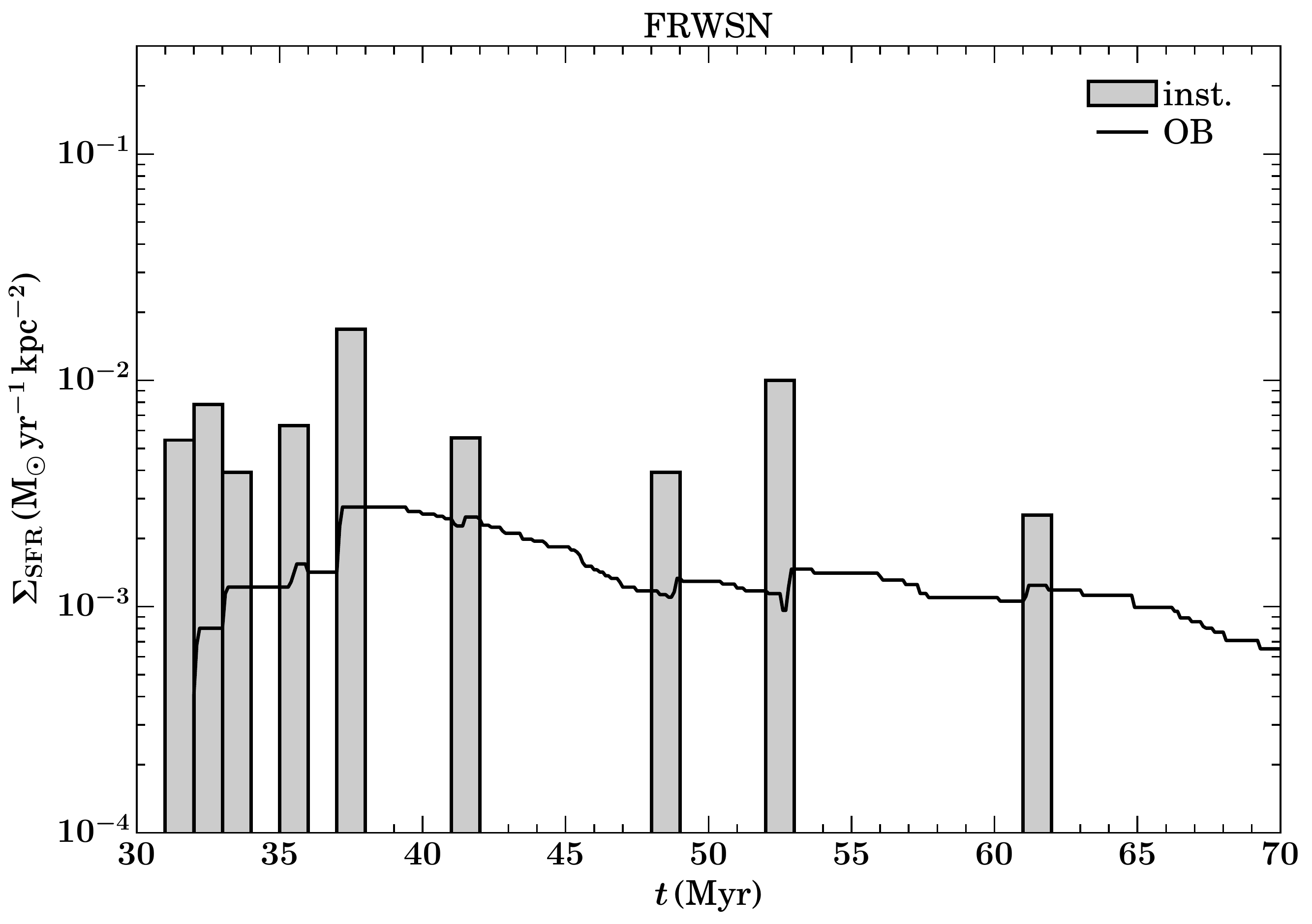}}
\caption{SFR surface density $\Sigma_\mathrm{SFR}$ as a function of time $t$ for runs FSN, FWSN and FRWSN (from top to bottom).
The panels show the instantaneous star formation events, binned over intervals of $1\,$Myr, and SFRs distributed over the OB
stars' lifetimes (solid lines).}
\label{fig:sfr}
\end{figure}

Since the ISM is shaped by stellar feedback, it is key for understanding the ISM in our simulations
to examine the stellar clusters that form.
Star formation takes place by formation of new sink
particles and by accretion onto already existing ones.
As explained in Section~\ref{sec:sim}, for each $120\,$M$_\odot$ of gas added to a sink particle,
we select a new massive star by sampling from a Salpeter IMF. 
Figure~\ref{fig:sfr} shows the SFR
surface density $\Sigma_\mathrm{SFR}$ as a function of time $t$ for all three simulations.
We compute the SFRs in two different ways (see also \citealt{gatto16}).
The instantaneous SFR is determined by summing the mass that is converted into
stars over time intervals of $1\,$Myr. It describes the individual star formation events
in the simulations.

However, when we want to compare the SFR in a simulation with
the SFR measured in synthetic H$\alpha$ images (Section~\ref{sec:halpha}), this is not
the SFR we expect to observe. Instead, an OB star emits ionising radiation during its entire
lifetime, not only when it has just formed. Therefore, we get a much better estimate
of the observed SFR when we distribute the $120\,$M$_\odot$ of newly formed stars over the
lifespan of the massive star associated with this cluster.
The SFRs computed in this way are shown as solid lines
in Figure~\ref{fig:sfr}.
One can clearly see how this SFR declines after star formation events.
This signifies the death of very massive stars ($M \geq 30\,$M$_\odot$) with short lifetimes
(less than $7\,$Myr). In contrast, a $9\,$M$_\odot$ star lives for $35\,$Myr before it explodes
as supernova. Because of the shape of the IMF, these stars at the lower end of the high-mass
slope of the IMF are the most abundant stars in the clusters. Therefore, the SFR always remains
positive at a significant level after the first star formation event, since these stars provide a floor to the observed SFR.

Comparing the three simulations in Figure~\ref{fig:sfr}, there are some notable differences.
In run FSN, star formation events occur steadily from the onset of star formation at $t = 30\,$Myr
until we stop the simulation at $t = 70\,$Myr. In contrast, runs FWSN and FRWSN display many fewer
star formation events. As a result, the averaged SFR in these simulations is reduced by one order
of magnitude compared to run FSN.

\begin{figure}
\centerline{\includegraphics[width=\linewidth]{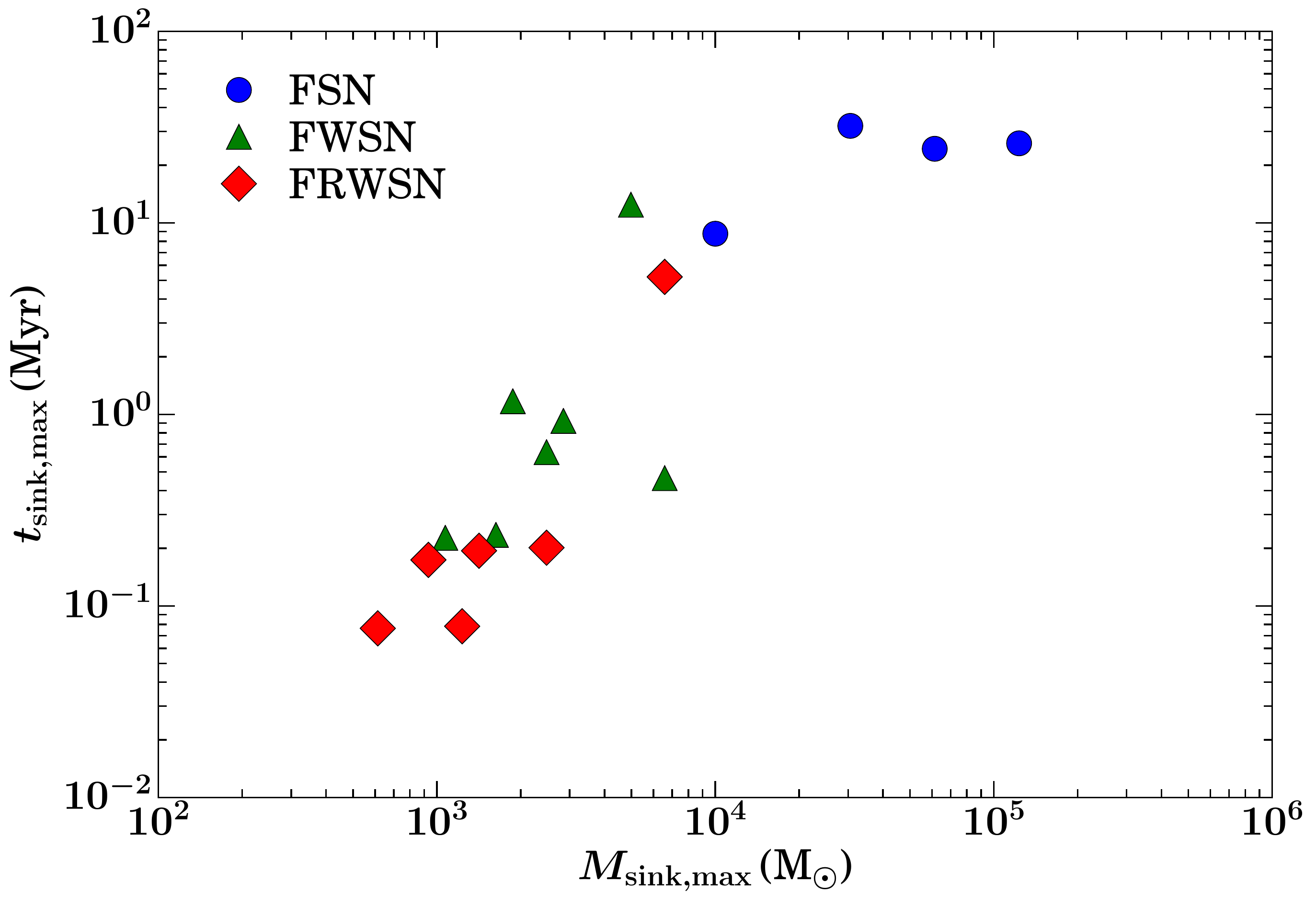}}
\caption{Time $t_\mathrm{sink,max}$ it takes for a sink particle to reach its final mass $M_\mathrm{sink,max}$ vs. $M_\mathrm{sink,max}$
for the three simulations.}
\label{fig:tvsm} 
\end{figure}

The reason for this behaviour is the self-regulation of star formation by early feedback (see also \citealt{gatto16}).
Figure~\ref{fig:tvsm} shows for each sink particle formed in the three simulations the time
$t_\mathrm{sink,max}$ that it takes to reach its final mass $M_\mathrm{sink,max}$ versus $M_\mathrm{sink,max}$.
In run FSN, the sink particles have final masses between $10^4$ and $10^5\,$M$_\odot$ and need some $10\,$Myr
to reach this mass. For the first few Myr after their formation, the sink particles in this simulation can accrete unimpededly
because no supernovae have exploded yet. In contrast, in run FWSN, stellar winds start blowing away the material close
to the sink particle immediately after its formation. As a consequence, the final sink particle masses are reduced
by one order of magnitude, and the typical timespan of sink particle accretion is reduced to only $0.1$ to $1\,$Myr.
In run FRWSN, where photoionisation feedback raises the thermal pressure by two orders of magnitude compared to the surrounding
molecular gas, accretion is stopped even more efficiently. Here, the majority of sink particles have masses around $10^3\,$M$_\odot$
and accrete for only $0.1\,$Myr.

It is important to note that both wind and radiation feedback are steady processes that commence with the
first star formation event and only cease when all stars in the cluster are gone. Therefore,
strong infall onto the sink particle is necessary to quench this feedback to the extent that accretion can continue for more
than $1\,$Myr, or to facilitate a second episode of star formation after the sink particle already contains a substantial
population of stars. In contrast, supernova feedback is highly intermittent. Therefore, in run FSN sink particles can
accrete even when supernovae are already exploding. In this case, accretion continues during the time intervals between
consecutive supernova explosions, when the gas has cooled sufficiently after a supernova injection.

\begin{figure}
\centerline{\includegraphics[width=\linewidth]{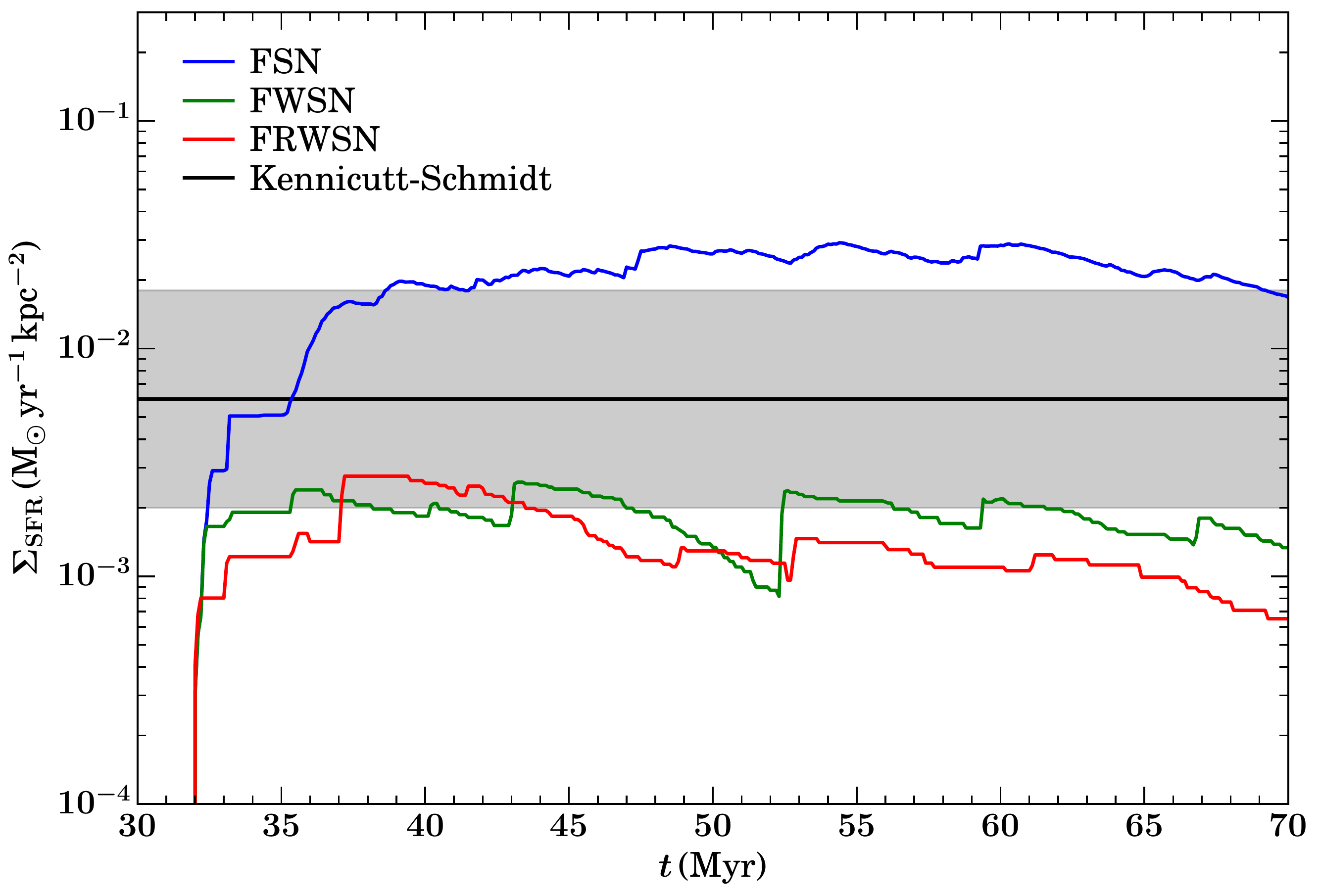}}
\caption{SFR surface density $\Sigma_\mathrm{SFR}$ as a function of time $t$ in the simulations.
The black line indicates the Kennicutt-Schmidt value, and the grey band a factor of three uncertainty.}
\label{fig:sfrOB}
\end{figure}

We can compare our SFR surface density to the observational data of \citet{leroy08}. For normal star-forming spiral galaxies
with $\Sigma_\mathrm{gas} = 10\,$M$_\odot\,$pc$^{-2}$, the average SFR surface density is around 
$\Sigma_\mathrm{SFR} = 6 \times 10^{-3}\,$M$_\odot\,$yr$^{-1}\,$kpc$^{-2}$, which corresponds to the Kennicutt-Schmidt value.
However, the data shows a significant scatter
around this number. Figure~\ref{fig:sfrOB} displays the SFR surface density $\Sigma_\mathrm{SFR}$ for the three simulations
as a function of time $t$. The plot also indicates a factor of three margin around the Kennicutt-Schmidt value,
which is within the scatter in the \citealt{leroy08} data for this $\Sigma_\mathrm{gas}$ (see also Figure~\ref{fig:dep}).
We find that run FSN is on the upper end of the margin, while runs FWSN and FRWSN are on the lower end.
The time-averaged SFR in run FRWSN is slightly lower than in run FWSN, indicating that early feedback by radiation is only
responsible for a small additional reduction on top of the already substantial lowering of the SFR by stellar winds.
Since we do not have a control run with radiation but without winds, we cannot say whether the winds are essential here,
or if radiation alone would have a comparable effect as the winds.

\begin{figure}
\centerline{\includegraphics[width=\linewidth]{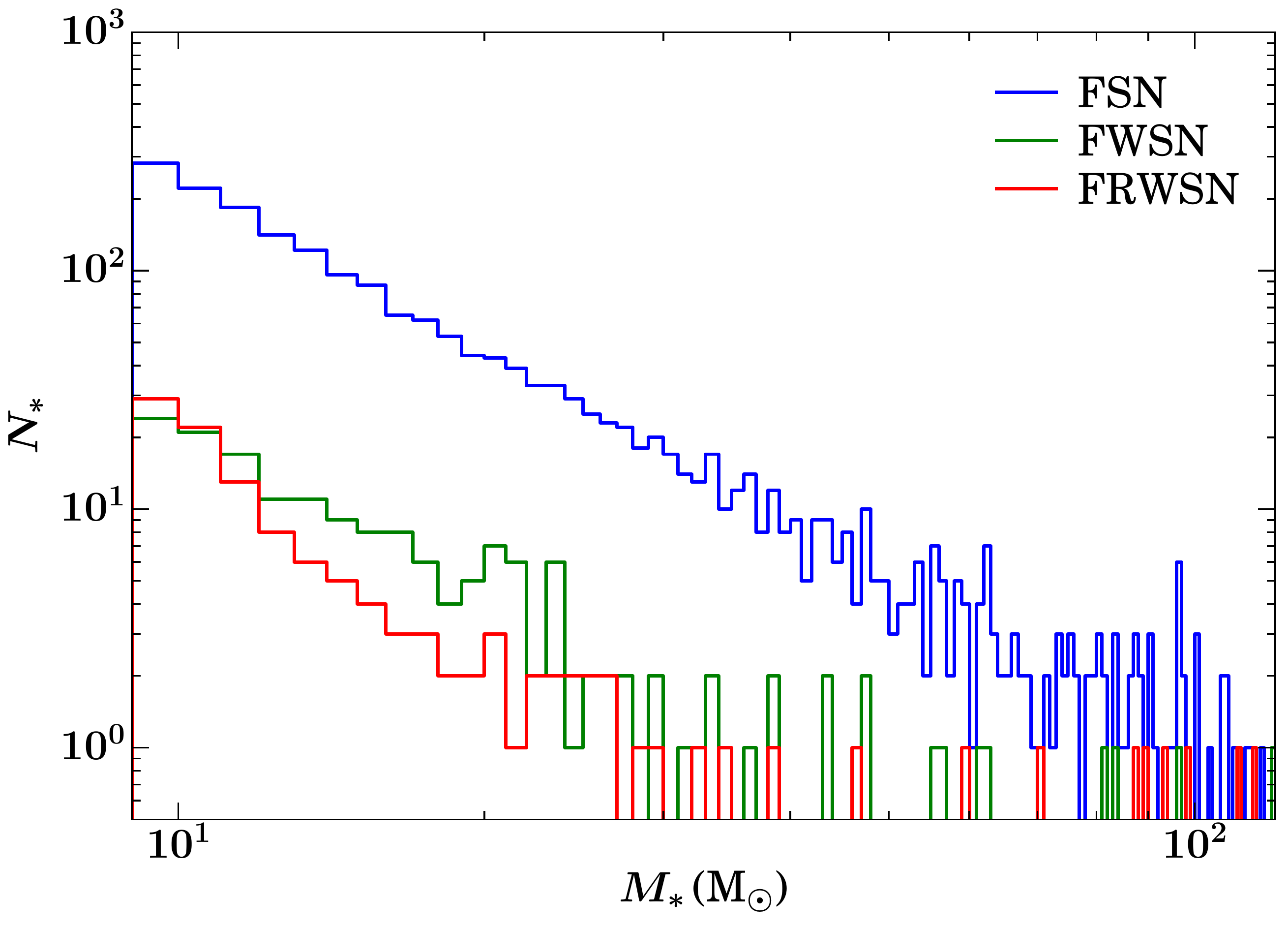}}
\caption{Histograms of all massive stars formed during the three simulations.}
\label{fig:hist}
\end{figure}

The reduced SFR in runs FWSN and FRWSN compared to run FSN has consequences for the stellar populations (star cluster sink particles) in the simulations.
Figure~\ref{fig:hist} shows histograms of all stars formed during the three runs. The number counts for run FSN are elevated by
a factor of $10$ compared to the other two simulations. In particular, while run FSN has a substantial population of very
massive stars with $M \geq 30\,$M$_\odot$, runs FWSN and FRWSN only have very few such stars. However, since both wind
and radiative energy output are a steep and highly non-linear function of stellar mass (compare Figure~\ref{fig:tracks}),
this small population still dominates the amount of energy injected into the ISM. We quantify this effect in Section~\ref{sec:enin}.

\section{Energy input}
\label{sec:enin}

\begin{figure*}
\centerline{\includegraphics[width=0.5\linewidth]{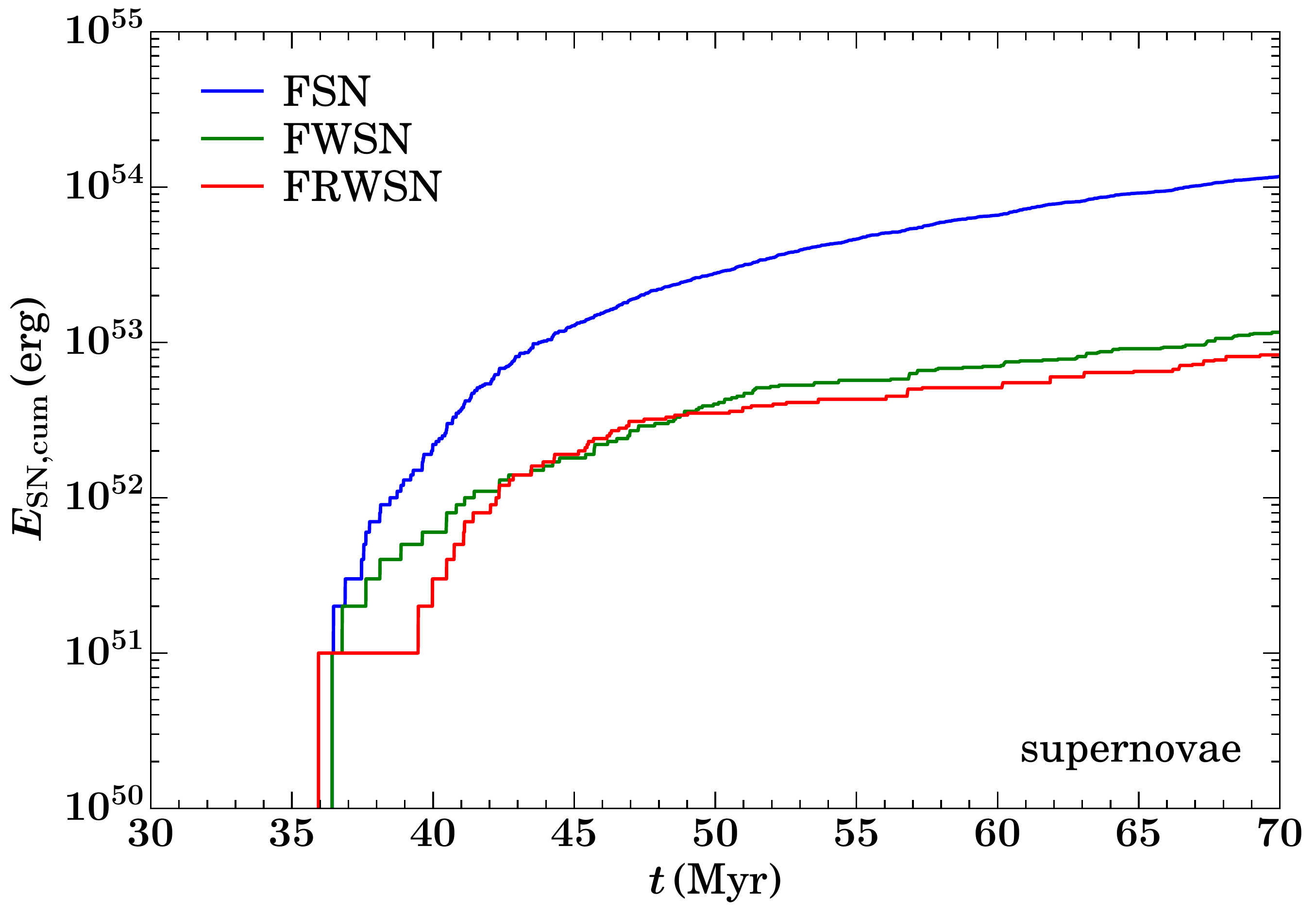}
\includegraphics[width=0.5\linewidth]{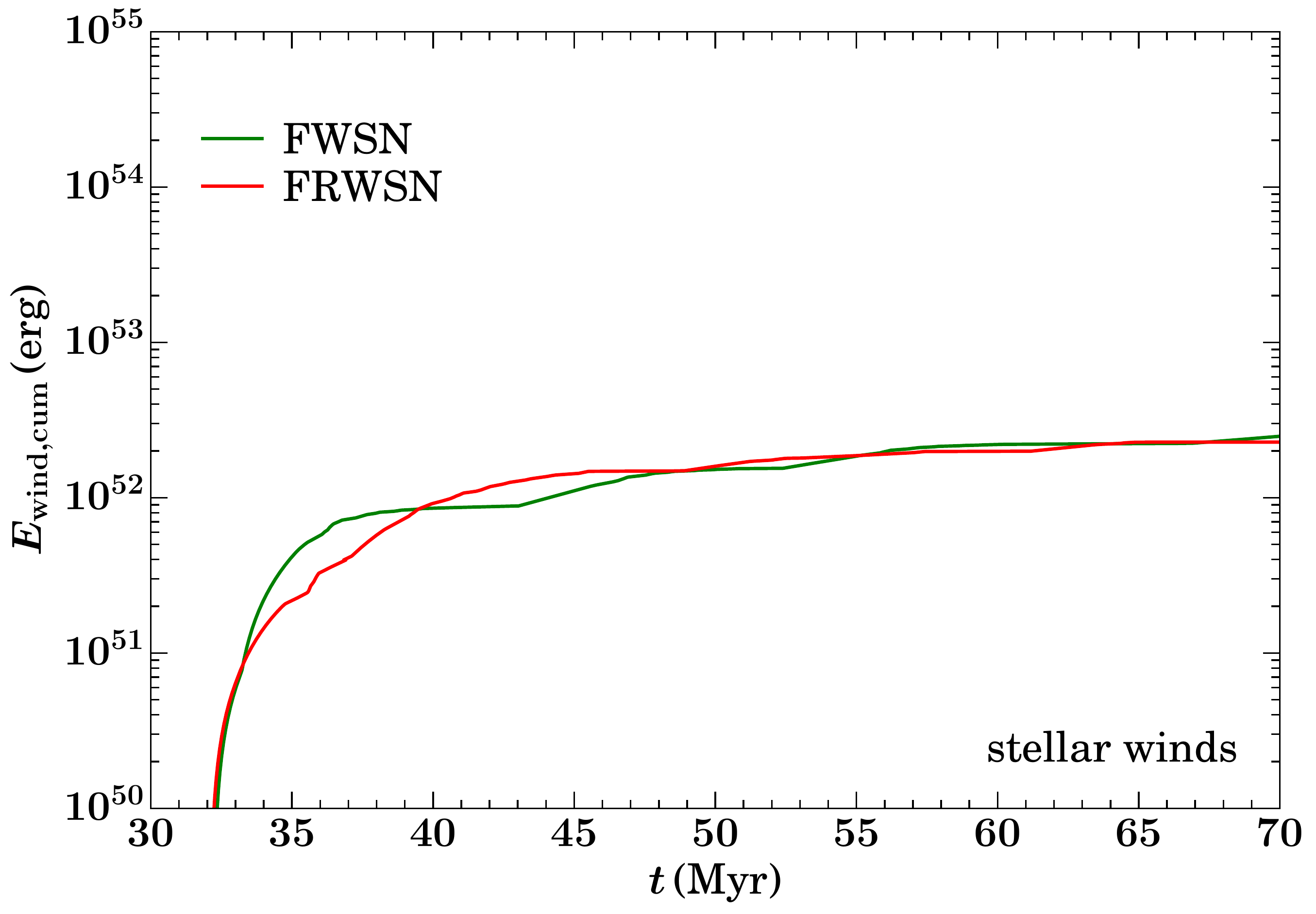}}
\caption{Cumulative energy injection by supernova explosions (left) and stellar winds (right) for the three simulations
as a function of time.
Runs FWSN and FRWSN have similar cluster sink properties and therefore similar energy injection statistics.
The corresponding plot for radiation is shown in Figure~\ref{fig:erad}.}
\label{fig:einjsnwinds}
\end{figure*}

We can study the impact of the different forms of stellar feedback on the ISM by considering the associated energies.
Figure~\ref{fig:einjsnwinds} shows the cumulative energy injected by supernova explosions $E_\mathrm{SN,cum}$ as a function of time $t$ for
the three simulations. Since run FSN forms roughly ten times more stars than the simulations with early feedback
FWSN and FRWSN, $E_\mathrm{SN,cum}$ is increased by a similar factor. As we have already seen, the SFR in runs FWSN
and FRWSN does not differ much, and therefore $E_\mathrm{SN,cum}$ is comparable in these simulations, too.

Figure~\ref{fig:einjsnwinds} also depicts the cumulative energy injected into the ISM by stellar winds $E_\mathrm{wind,cum}$. Here also, the simulations
FWSN and FRWSN behave very similarly at all times. In both runs, $E_\mathrm{wind,cum}$ at the end of the simulation $t = 70\,$Myr is
around $2 \times 10^{52}\,$erg, which should be compared to $E_\mathrm{SN,cum}$ at that time, which is $10^{53}\,$erg. 

\begin{figure}
\centerline{\includegraphics[width=\linewidth]{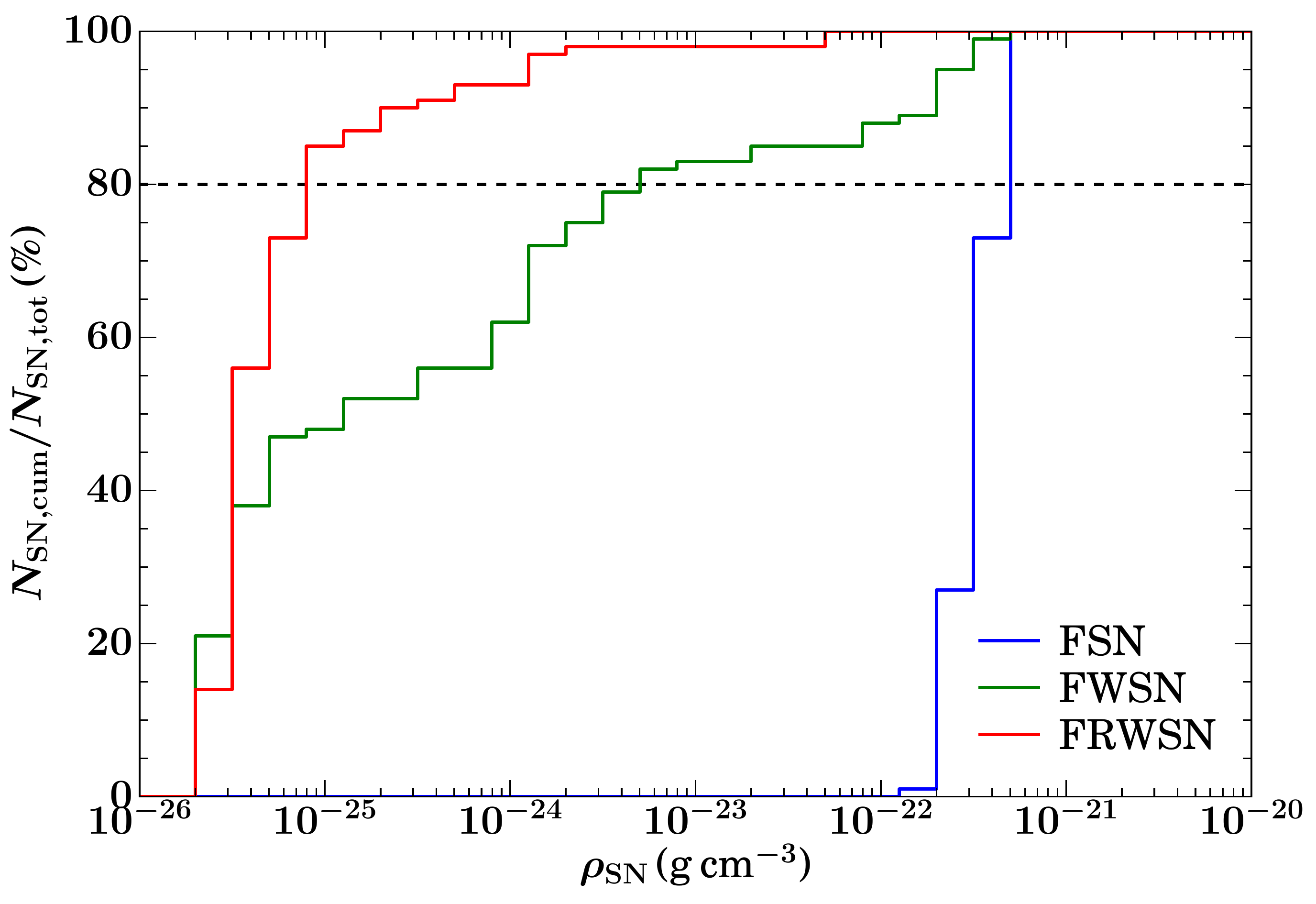}}
\caption{Normalised cumulative distribution of supernovae as a function of the mean environmental density in which they explode
for the three simulations.
The inclusion of winds reduces the ambient density of $80\,$\% of the supernova explosions from less than $\sim10^{-22}$\,g\,cm$^{-3}$
to less than $\sim10^{-23}$\,g\,cm$^{-3}$, which is further dramatically reduced to less than $\sim10^{-25}$\,g\,cm$^{-3}$ by radiation.
}
\label{fig:sndist}
\end{figure}

The impact of a supernova injection depends on the ambient density of the gas. Since we inject thermal energy with each supernova,
the effect of an explosion grows with decreasing ambient density.
Dense gas cools radiatively very quickly, whereas underdense
gas stays hot for a long time. Figure~\ref{fig:sndist} shows the cumulative distribution of supernovae as a function of
the mean environmental density $\rho_\mathrm{SN}$ in which they explode, normalised to the total number of supernova explosions.
In run FSN, essentially all supernovae go off at a mean density around $\rho_\mathrm{SN} = 3 \times 10^{-22}\,$g\,cm$^{-3}$,
which is two orders of magnitude lower than the threshold density for sink particle formation $\rho_\mathrm{thr}$.
In simulation FWSN, where stellar winds blow the dense gas out of the supernova injection region, the distribution function
becomes much broader. The lowest mean density reached is only $\rho_\mathrm{SN} = 2 \times 10^{-26}\,$g\,cm$^{-3}$,
and the median of the distribution is $\rho_\mathrm{SN} = 10^{-25}\,$g\,cm$^{-3}$. Therefore, half of all supernovae
explode in very low-density gas. In run FRWSN, the distribution is similarly broad, but the median is only $\rho_\mathrm{SN} = 3 \times 10^{-26}\,$g\,cm$^{-3}$
in this case. This is the effect of the additional photoevaporation flow in this simulation. Therefore,
run FRWSN creates the largest amount of hot gas per supernova, followed by run FWSN and run FSN.
On the other hand, run FRWSN has the smallest SFR.

\begin{figure*}
\centerline{\includegraphics[width=0.5\linewidth]{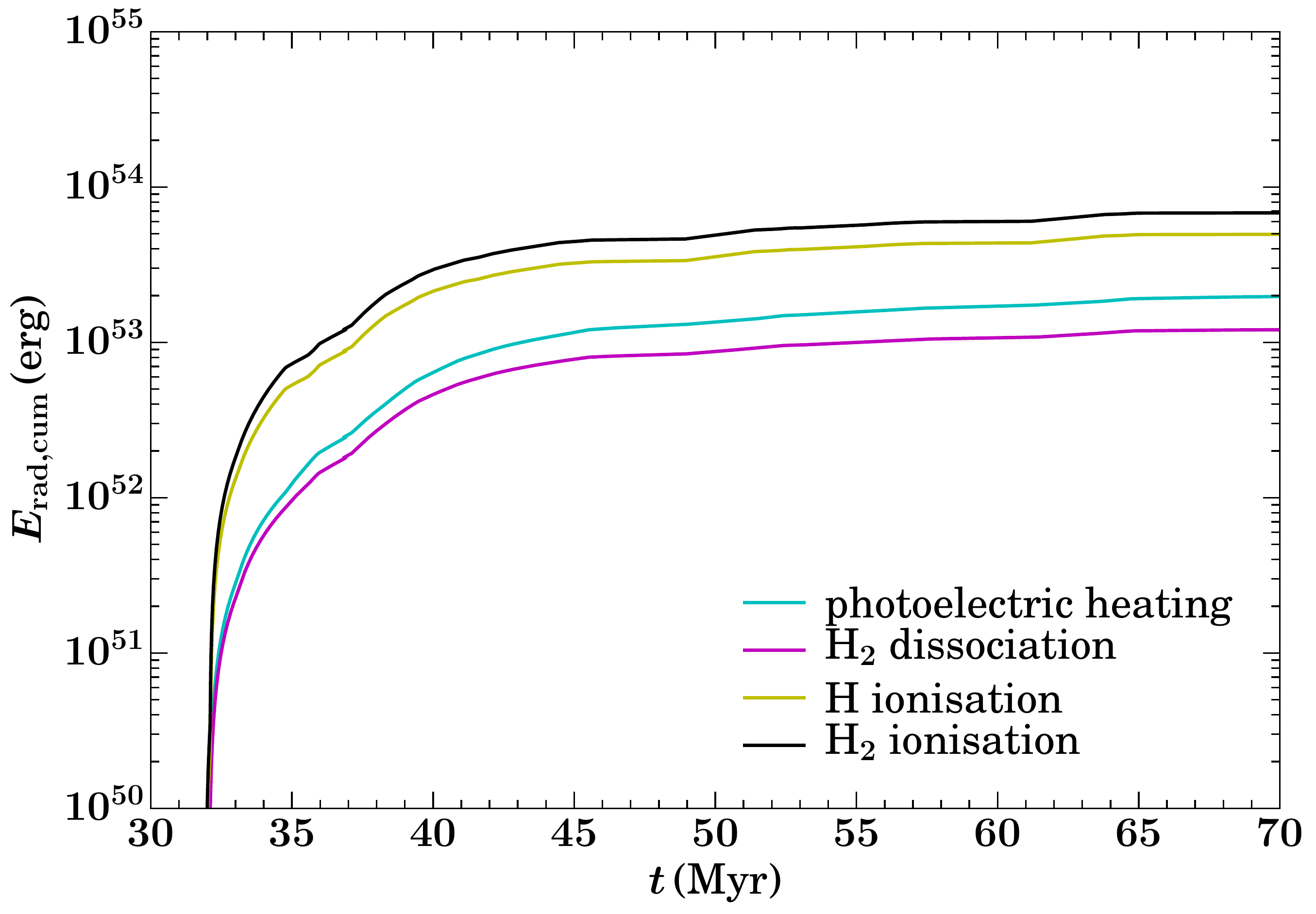}
\includegraphics[width=0.5\linewidth]{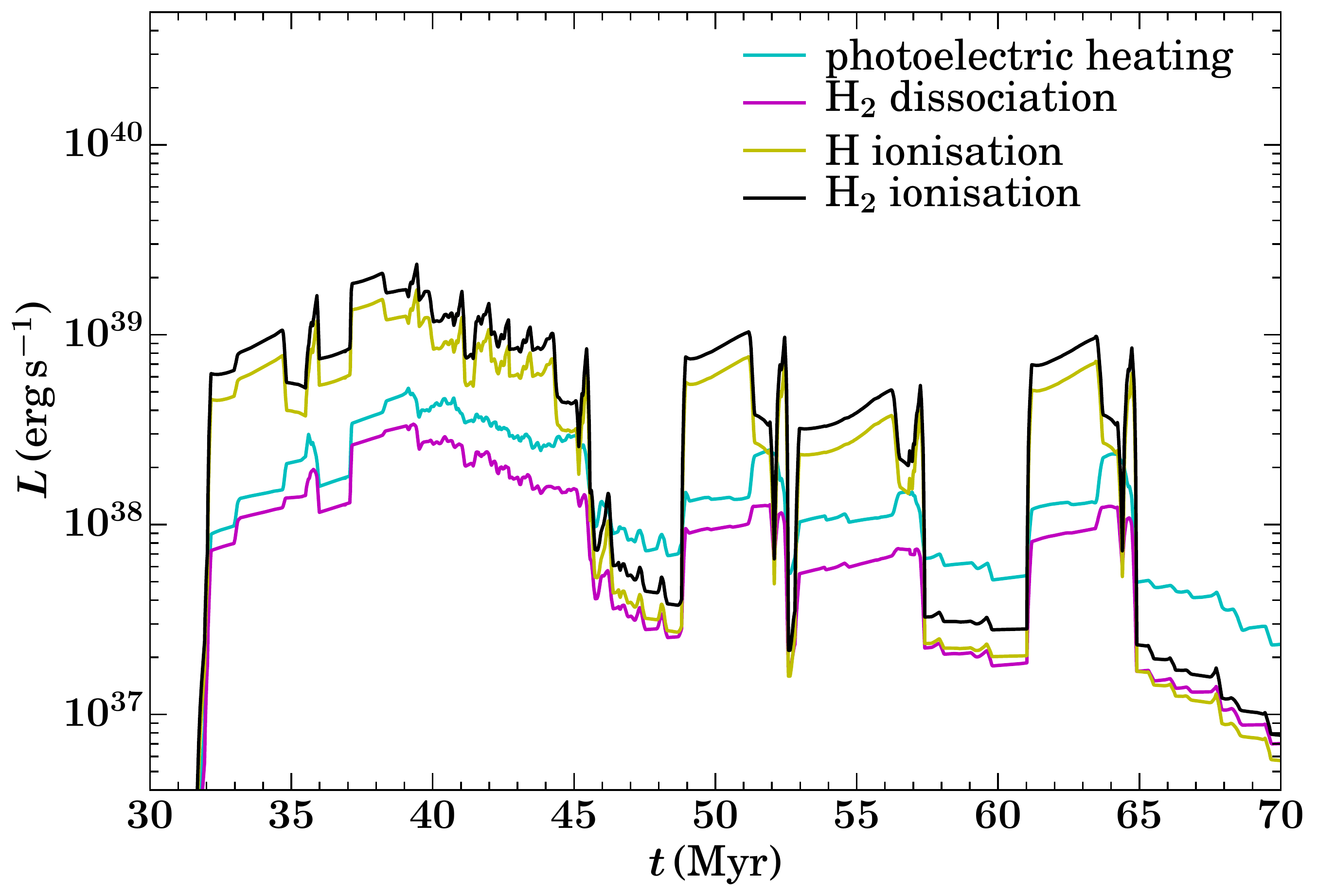}}
\caption{Left: Cumulative radiation energy output in the different photochemical processes of all stellar clusters in run FRWSN.
Right: Instantaneous luminosity as a function of time.
The total energy output is dominated by photoionisation of H$_2$ and H, followed by photoelectric heating and H$_2$ dissociation.
}
\label{fig:erad}
\end{figure*}

The energy input by winds and supernovae should be compared with the energy released by the star clusters in the form of radiation
in run FRWSN.
Figure~\ref{fig:erad} displays the cumulative energy available to the different photochemical processes included in \texttt{Fervent},
computed in the same way as in Section~\ref{sec:bud}.
Already H$_2$ dissociation alone can impart more energy to the ISM than the supernova explosions, provided that this radiation is
actually absorbed. The largest amount of energy can be transferred to the ISM via photoionisation heating, with a total amount of
$1.2 \times 10^{54}\,$erg at $t = 70\,$Myr with H$_2$ and H photoionisation combined.
Of course, neither photoelectric heating nor photoionisation or photodissociation can produce
$10^6\,$K hot gas like winds or supernovae. But considering the total energy budget in the ISM, the contribution from radiation
is very substantial. Even if only one tenth of all available radiation energy can be tapped by the material, this energy still
exceeds the combined energy input by winds and supernovae.

In Figure~\ref{fig:erad}, we also show the instantaneous luminosity of the different processes as a function of time.
The curves show several jumps where the radiative output suddenly drops by orders of magnitude. The origin of this variability
is the strong dependence of the luminosity on stellar mass (compare Figure~\ref{fig:tracks}). As discussed in Section~\ref{sec:stcl},
the star clusters of run FRWSN contain only very few stars with $M \geq 30\,$M$_\odot$ (see Figure~\ref{fig:hist}). These stars
dominate the cluster luminosity as long as they are present, but they live only for a few Myr. After they have exploded as supernovae,
only less massive stars survive that produce a lower luminosity.

\begin{figure*}
\centerline{\includegraphics[width=0.5\linewidth]{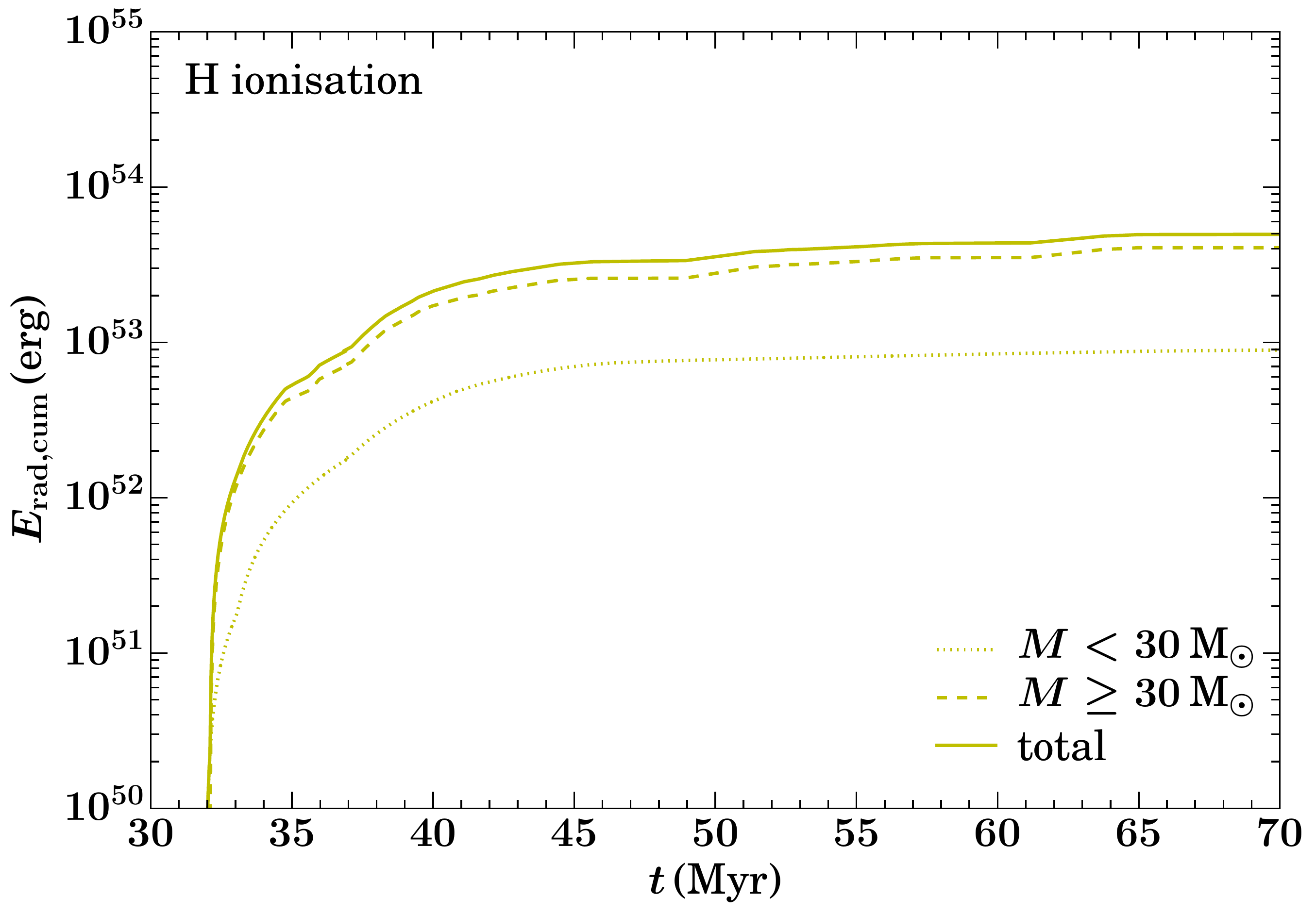}
\includegraphics[width=0.5\linewidth]{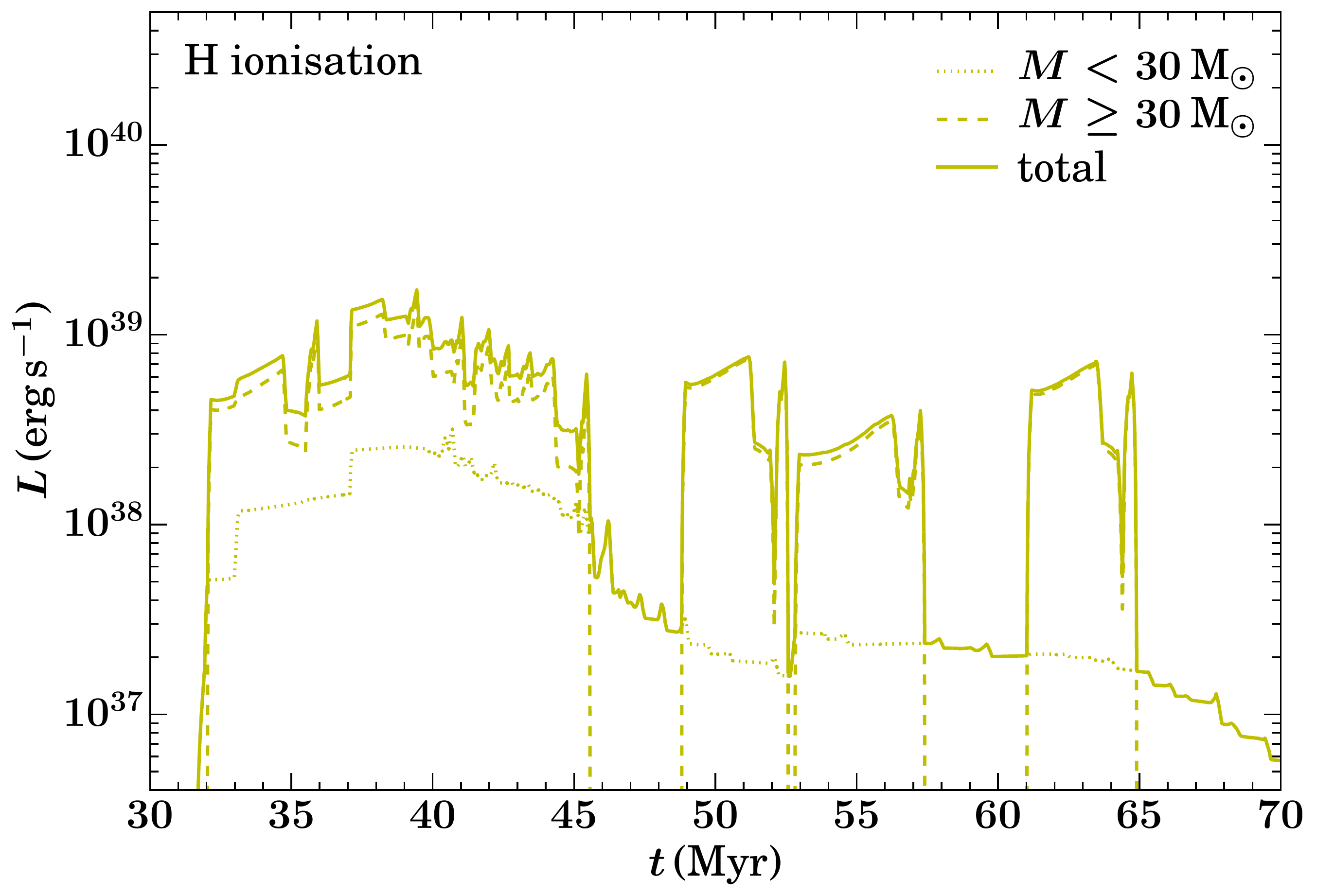}}
\caption{Same as Figure~\ref{fig:erad}, but for atomic hydrogen ionisation only. We have separated the contributions from
stars with masses $M < 30\,$M$_\odot$ (dotted) and $M \geq 30\,$M$_\odot$ (dashed) to the total (solid).}
\label{fig:eradbin}
\end{figure*}

Figure~\ref{fig:eradbin} illustrates this effect. It shows the radiative energy output associated with the ionisation of atomic hydrogen
as a function of time. We plot the energy emitted by less massive stars with $M < 30\,$M$_\odot$, very massive stars with
$M \geq 30\,$M$_\odot$, and the sum of the two. The cumulative plot shows that the former contribute only $20\,$\% to the total
radiation energy. The plot of the instantaneous luminosity emitted by the clusters reveals that
very massive stars with $M \geq 30\,$M$_\odot$ dominate the radiative output by an order of magnitude or more. When these
stars disappear, the curve drops to a floor value produced by the less massive stars with $M < 30\,$M$_\odot$
\citep[see e.g.][for a detailed analysis of the effect of stochastic stellar populations on estimated SFRs]{dasilva14,krum15}.

\section{Mass fractions}
\label{sec:mf}
\begin{figure*}
\centerline{\includegraphics[width=0.5\linewidth]{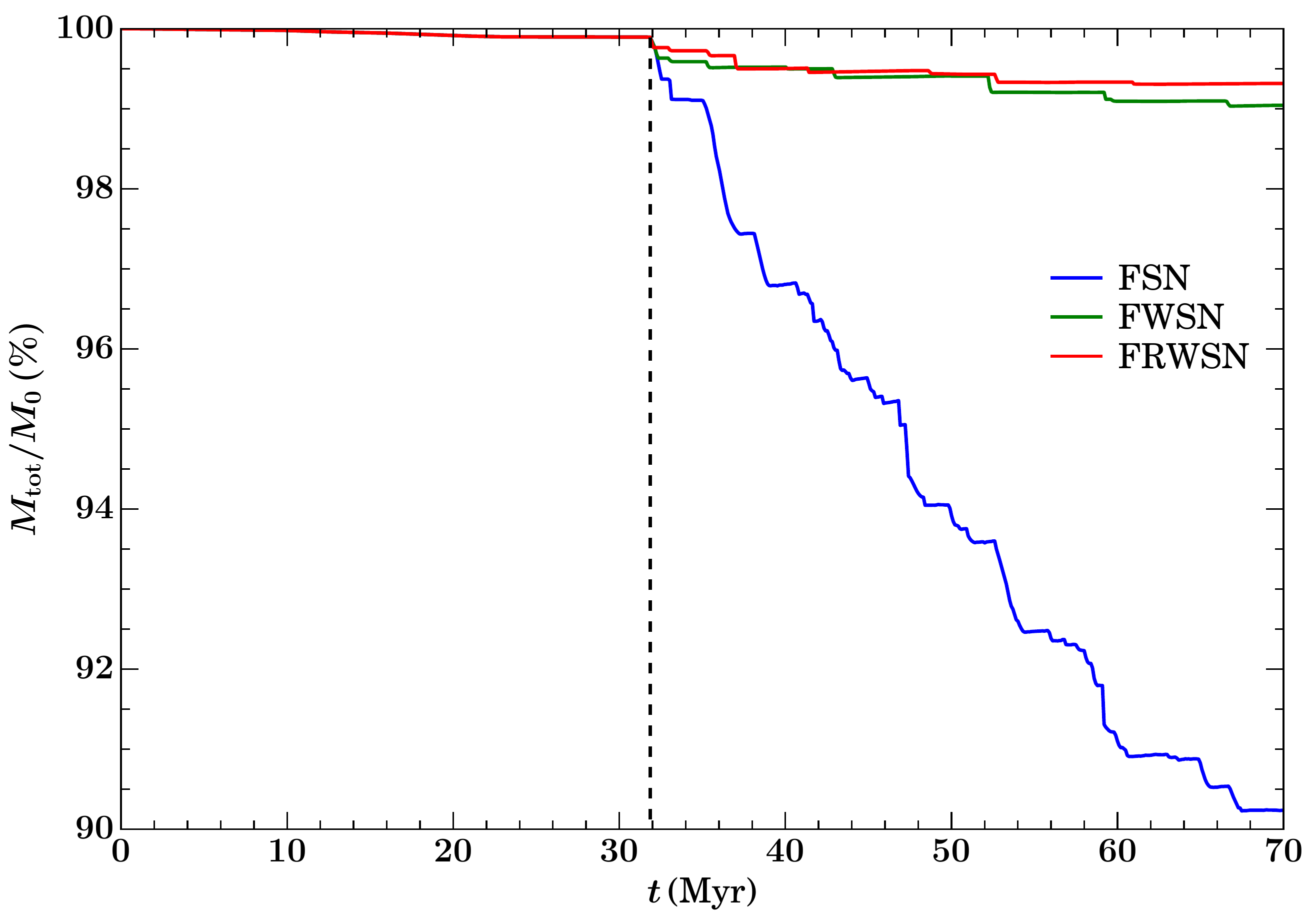}
\includegraphics[width=0.5\linewidth]{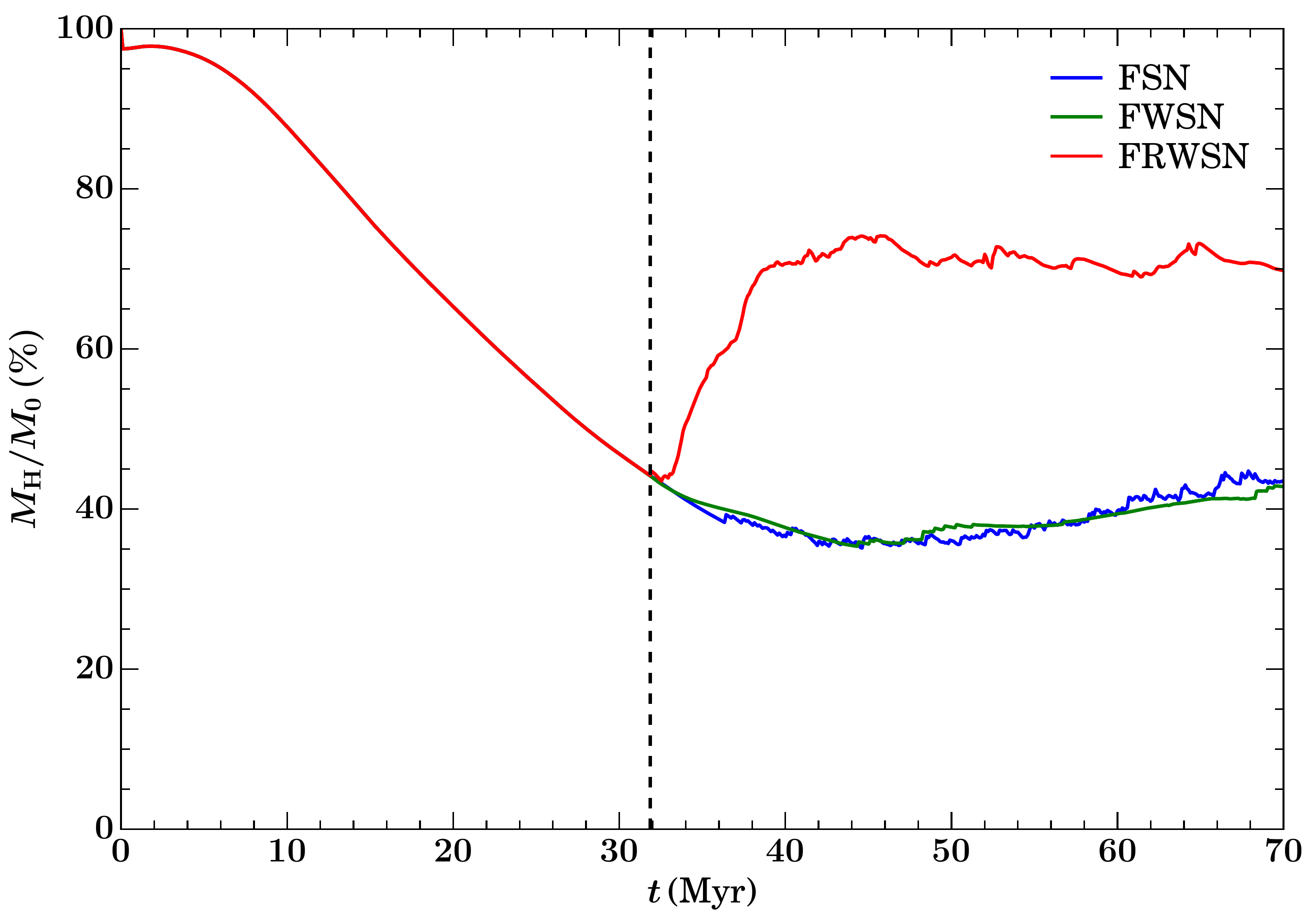}}
\centerline{\includegraphics[width=0.5\linewidth]{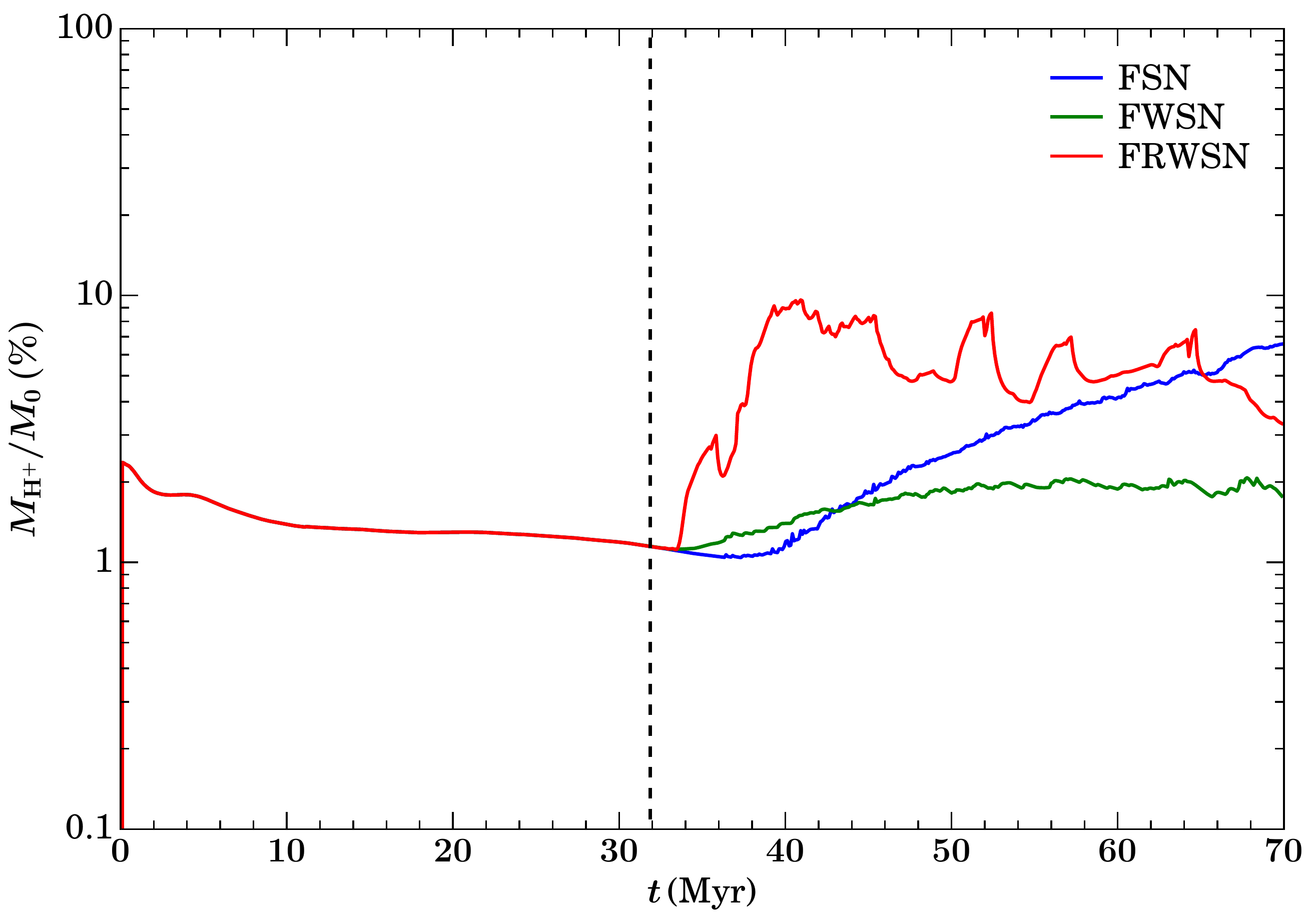}
\includegraphics[width=0.5\linewidth]{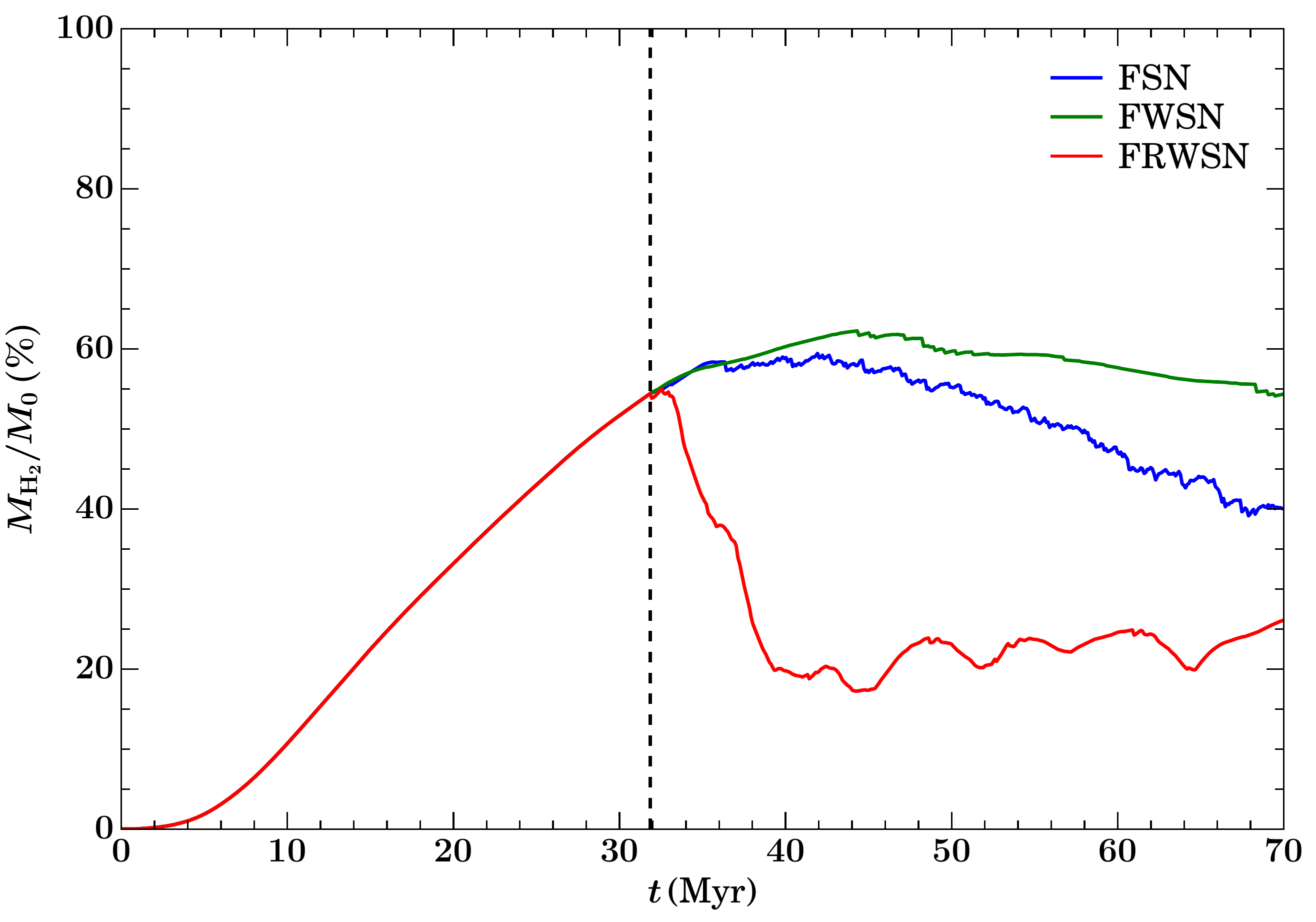}}
\caption{Total mass and mass fraction of atomic, ionised and molecular hydrogen as a function
of time $t$ for all simulations (from top left to bottom right). All masses are normalised to $M_0$,
the total mass at $t = 0$. The drop of the total mass over time is due to accretion of gas by sink particles.
The vertical dashed line marks the onset of star formation.}
\label{fig:mf}
\end{figure*}

The different stellar populations and energy input in the three simulations are reflected in the resulting ISM.
Figure~\ref{fig:mf} shows the time evolution of the total mass and of the mass fractions of atomic, ionised and
molecular hydrogen. In run FSN, about $10\,$\% of the initial gas mass gets accreted onto sink particles
during the simulation. For run FWSN, it is only $1\,$\%, and for run FRWSN even less. These results are
consistent with the trend in the SFRs for the three simulations (compare Figure~\ref{fig:sfrOB}).

Interestingly, the evolution of the atomic hydrogen mass fraction is almost identical in runs FSN
and FWSN, with a value around $40\,$\%. Simulation FRWSN, however, has a mass fraction of atomic hydrogen
around $70\,$\%. This is due to the effect of photodissociation, which converts  molecular into
atomic gas.

The mass fraction of ionised hydrogen starts to grow monotonically $10\,$Myr after the onset of star formation
from $1\,$\% to $7\,$\% in run
FSN. This is the result of the supernova explosions, which produce an increasing amount of ionised gas.
Run FWSN has a roughly constant ionised hydrogen mass fraction of $2\,$\%. Due to the reduced SFR in this
simulation, many fewer supernovae explode compared to run FSN, and the additional collisional ionisation by stellar wind feedback
cannot compensate for this effect. The ionised hydrogen mass fraction of run FRWSN oscillates between
$3\,$\% and $10\,$\% as a function of time. These oscillations are very closely correlated with the
variability in the radiative output from the star clusters (compare Figure~\ref{fig:erad}).
This clearly demonstrates that photoionisation from stellar radiation is the primary source of ionised gas
in this simulation, with supernovae and winds contributing only a negligible amount.

The molecular hydrogen mass fraction of run FSN drops from $60\,$\% to $40\,$\% from $40\,$Myr onwards. The delay of $10\,$Myr
between the onset of star formation and the reduction of the molecular hydrogen mass fraction suggests that it is due to supernova feedback,
since the evolution of the atomic and ionised hydrogen mass fractions show a similar behaviour. In contrast,
if this reduction was primarily due to accretion of molecular gas onto the sink particles, one would not expect such
a delay. For run FWSN, the molecular hydrogen mass fraction stays at around $60\,$\%. In run FRWSN,
the molecular hydrogen mass fraction oscillates around a value of $20\,$\%. These oscillations are
again indicative of photoevaporation processes caused by the stellar irradiation of the molecular clouds.

\section{Depletion time}
\label{sec:dep}

\begin{figure}
\centerline{\includegraphics[width=\linewidth]{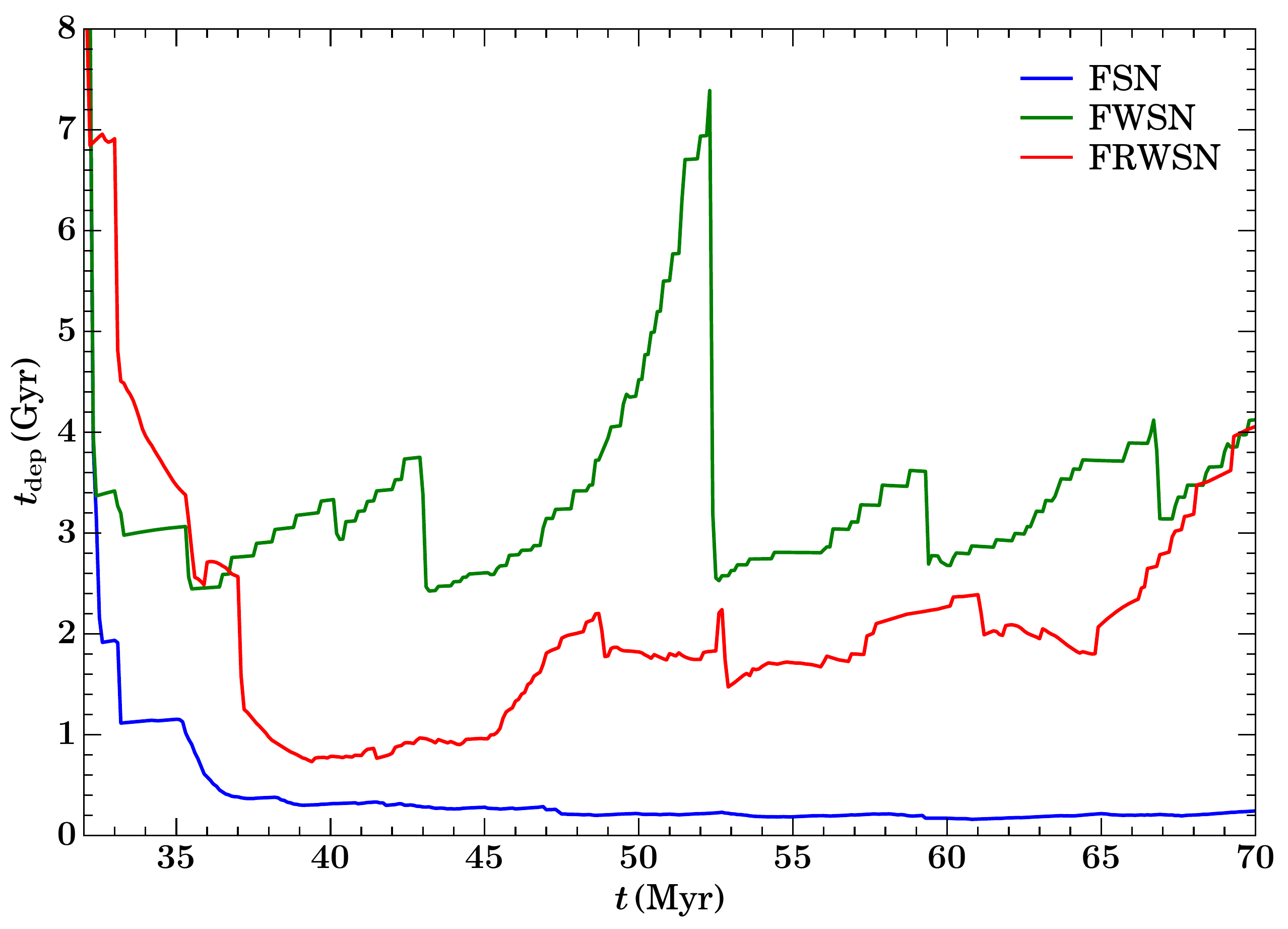}}
\centerline{\includegraphics[width=\linewidth]{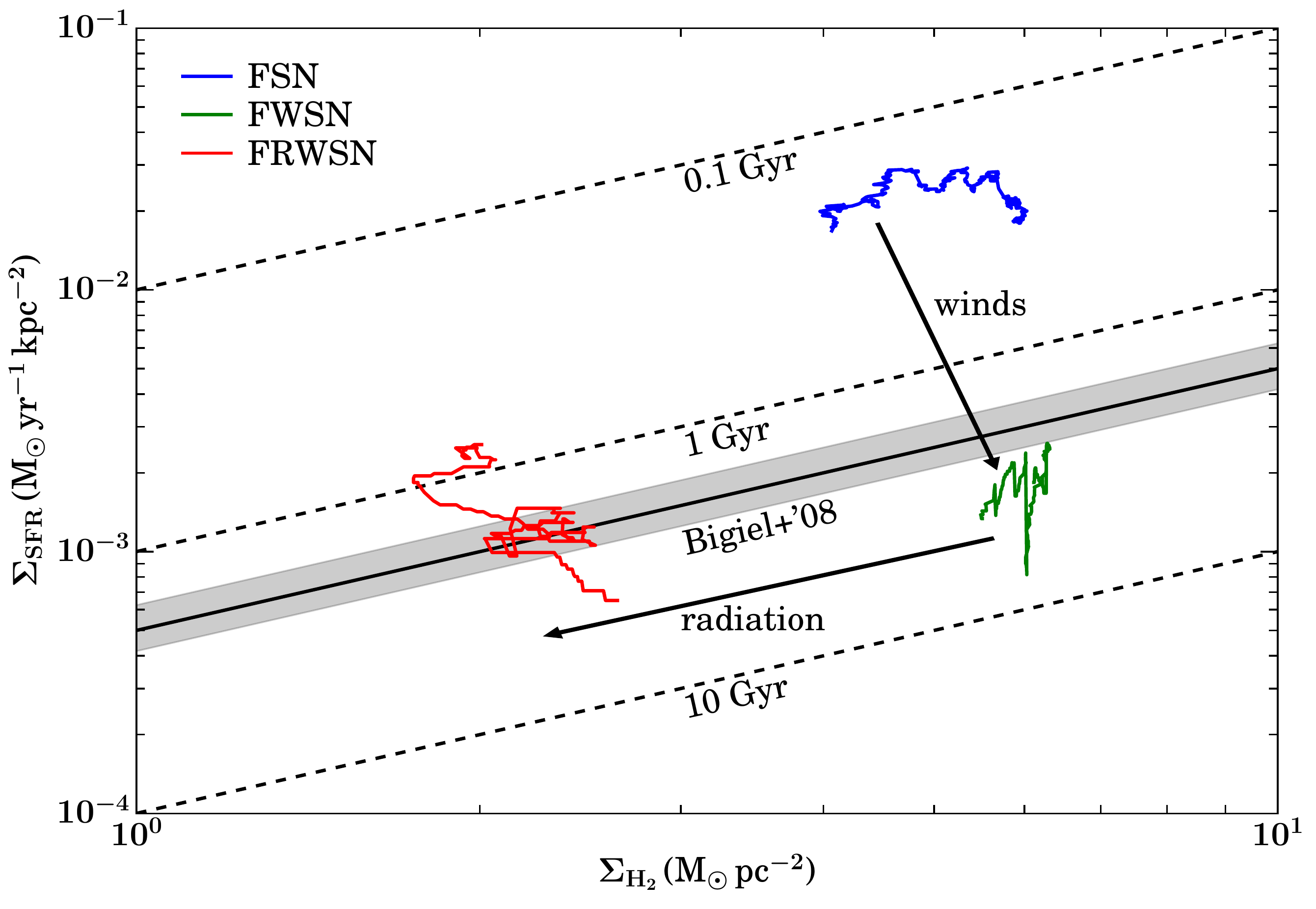}}
\centerline{\includegraphics[width=\linewidth]{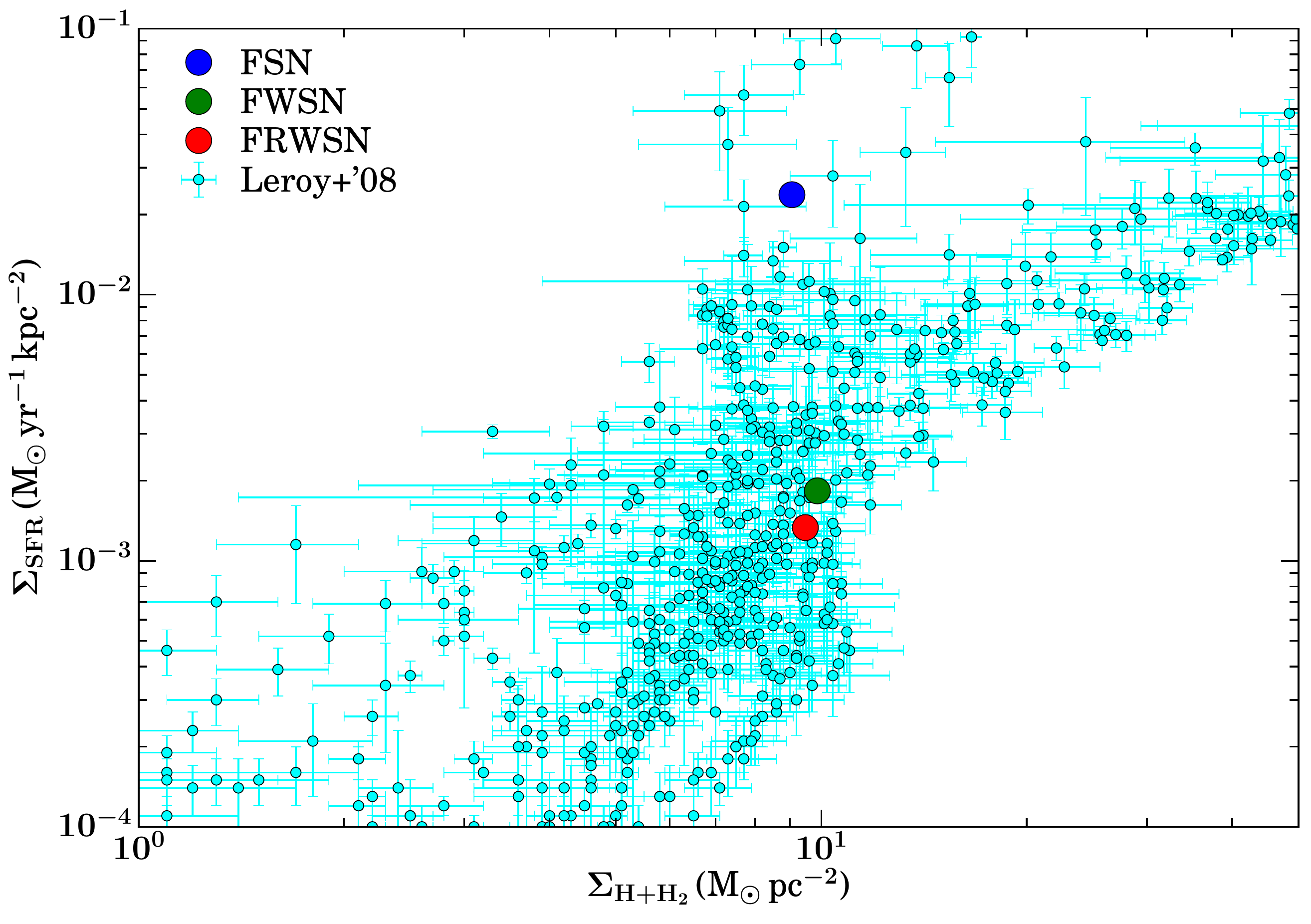}}
\caption{Top: Depletion time $t_\mathrm{dep} = \Sigma_{\mathrm{H}_2} / \Sigma_{\mathrm{SFR}}$ for the three simulations
as a function of time $t$. Middle: Plot of $\Sigma_{\mathrm{SFR}}$ versus $\Sigma_{\mathrm{H}_2}$ for the three simulations
for all snaphots after $t = 40\,$Myr. The dashed lines show constant depletion times of $0.1$, $1$ and $10\,$Gyr (from top
to bottom). The solid line represents the average value $t_\mathrm{dep} = 2\,$Gyr from \citet{bigiel08}, and the grey
band marks the observational scatter of $0.8\,$Gyr in their data.
Bottom: Plot of $\Sigma_{\mathrm{SFR}}$ and $\Sigma_{\mathrm{H}+\mathrm{H}_2}$ time-averaged from $t = 40\,$Myr to $t = 70\,$Myr
for the three simulations together with the data from \citet{leroy08}.}
\label{fig:dep}
\end{figure}

Since we dynamically form H$_2$ in the simulations, it is interesting to check whether the relation
between the molecular gas surface density $\Sigma_{\mathrm{H}_2}$ and the SFR surface density $\Sigma_{\mathrm{SFR}}$
obtained in the simulations is in agreement with observations. The ratio of these two quantities is the depletion
time $t_\mathrm{dep} = \Sigma_{\mathrm{H}_2} / \Sigma_{\mathrm{SFR}}$. 
Using $\Sigma_{\mathrm{SFR}}$ from Figure~\ref{fig:sfrOB} and the instantaneous $\Sigma_{\mathrm{H}_2}$ measured
in face-on projections of the disc, we can plot $t_\mathrm{dep}$ as a function of time for the three simulations.
The result is shown in Figure~\ref{fig:dep}.

In the data analysed by \citet{bigiel08}, the average depletion time is $t_\mathrm{dep} = 2\,$Gyr. Only the
run FRWSN reproduces this value. The SFR in run FSN is too high for the amount of molecular gas present,
and run FWSN has too much molecular gas for such a small SFR. Compared to run FWSN, radiation reduces $\Sigma_{\mathrm{H}_2}$ more
than $\Sigma_{\mathrm{SFR}}$, and $t_\mathrm{dep}$ falls on the observed value.

Figure~\ref{fig:dep} also shows $\Sigma_{\mathrm{SFR}}$ versus $\Sigma_{\mathrm{H}_2}$ for the three simulations.
We have omitted the data prior to $t = 40\,$Myr, because this is the time required for the H$_2$ mass fraction
in run FRWSN to converge and become relatively constant (compare Figure~\ref{fig:mf}). For most of the time, run FRWSN is within
the observational scatter of $0.8\,$Gyr around the average depletion time in the data of \citet{bigiel08}.
The curve for run FRWSN in this plot nicely illustrates the effect of star formation self-regulation.

It is also interesting to consider the relation between $\Sigma_{\mathrm{SFR}}$ and $\Sigma_{\mathrm{H}+\mathrm{H}_2}$
instead of $\Sigma_{\mathrm{H}_2}$. In Figure~\ref{fig:dep} we plot the data from \citet{leroy08} together with
the time-averaged quantities from $t = 40\,$Myr to $t = 70\,$Myr for the three simulations.
As already remarked in Section~\ref{sec:stcl}, our values are within the scatter of the observational data.
Run FSN is near the upper end in $\Sigma_{\mathrm{SFR}}$, but runs FWSN and FRWSN are in the middle of the scatter
for their $\Sigma_{\mathrm{H}+\mathrm{H}_2}$.

\section{Volume filling fractions}
\label{sec:vff}

\begin{figure*}
\centerline{\includegraphics[width=0.5\linewidth]{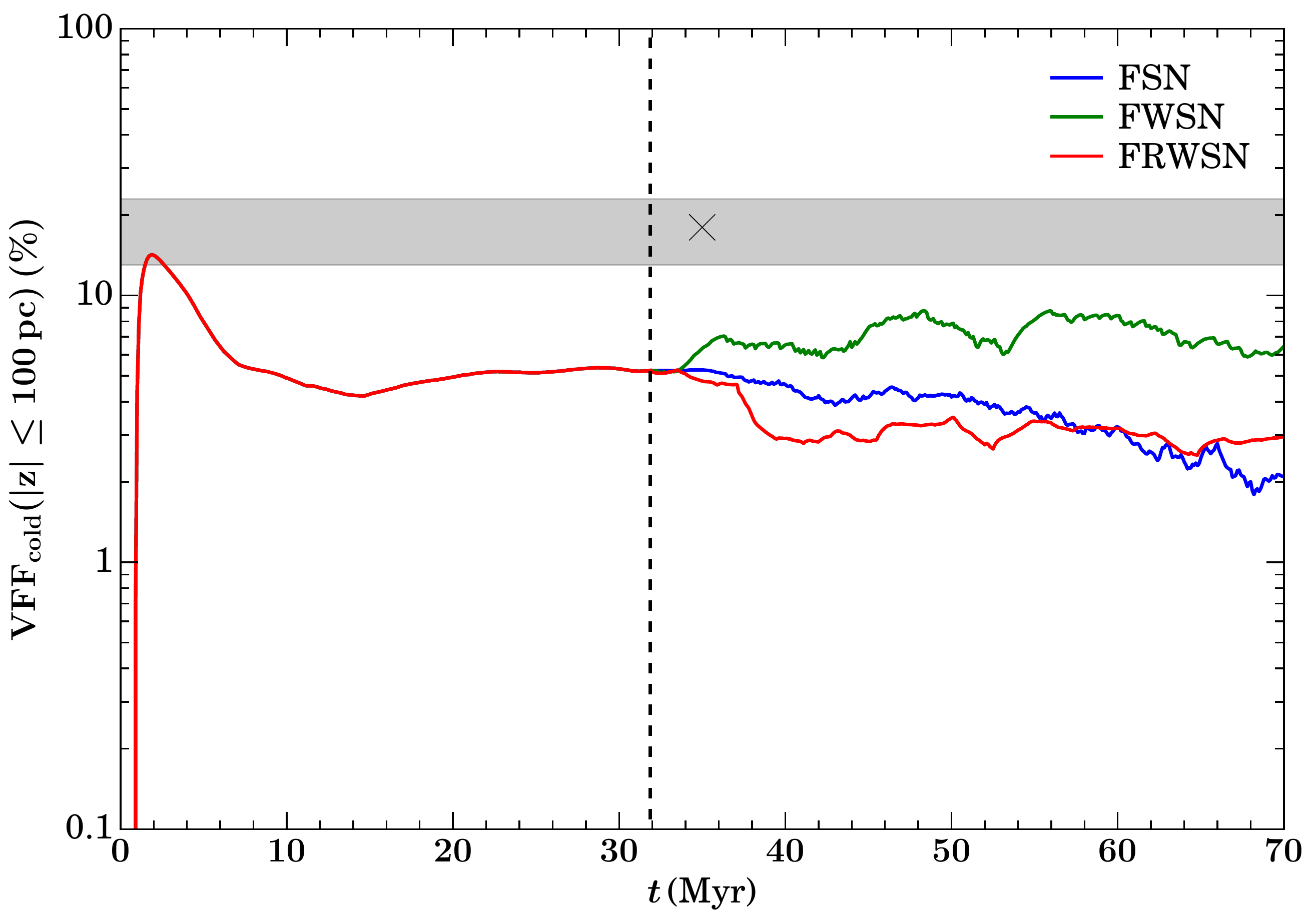}
\includegraphics[width=0.5\linewidth]{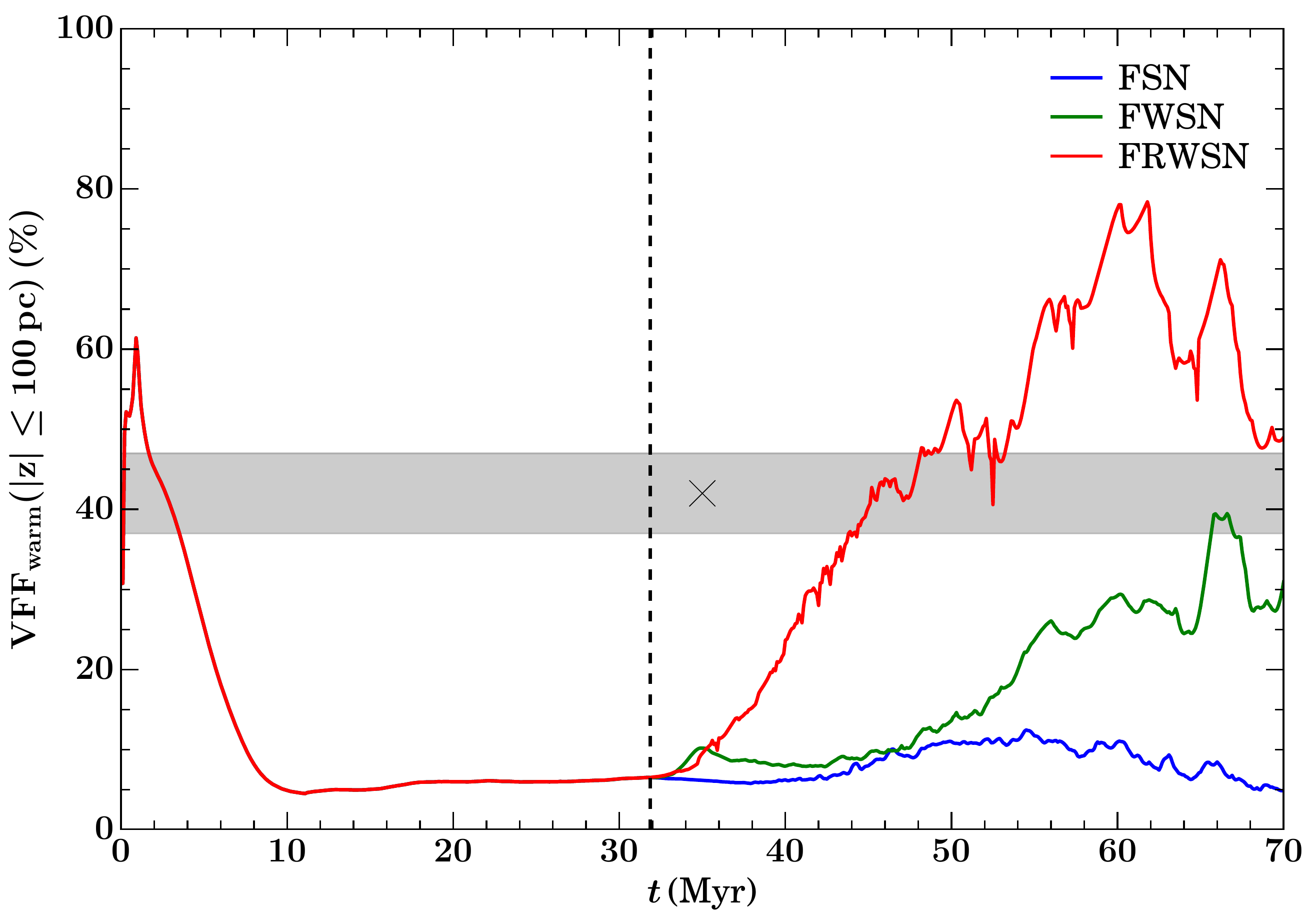}}
\centerline{\includegraphics[width=0.5\linewidth]{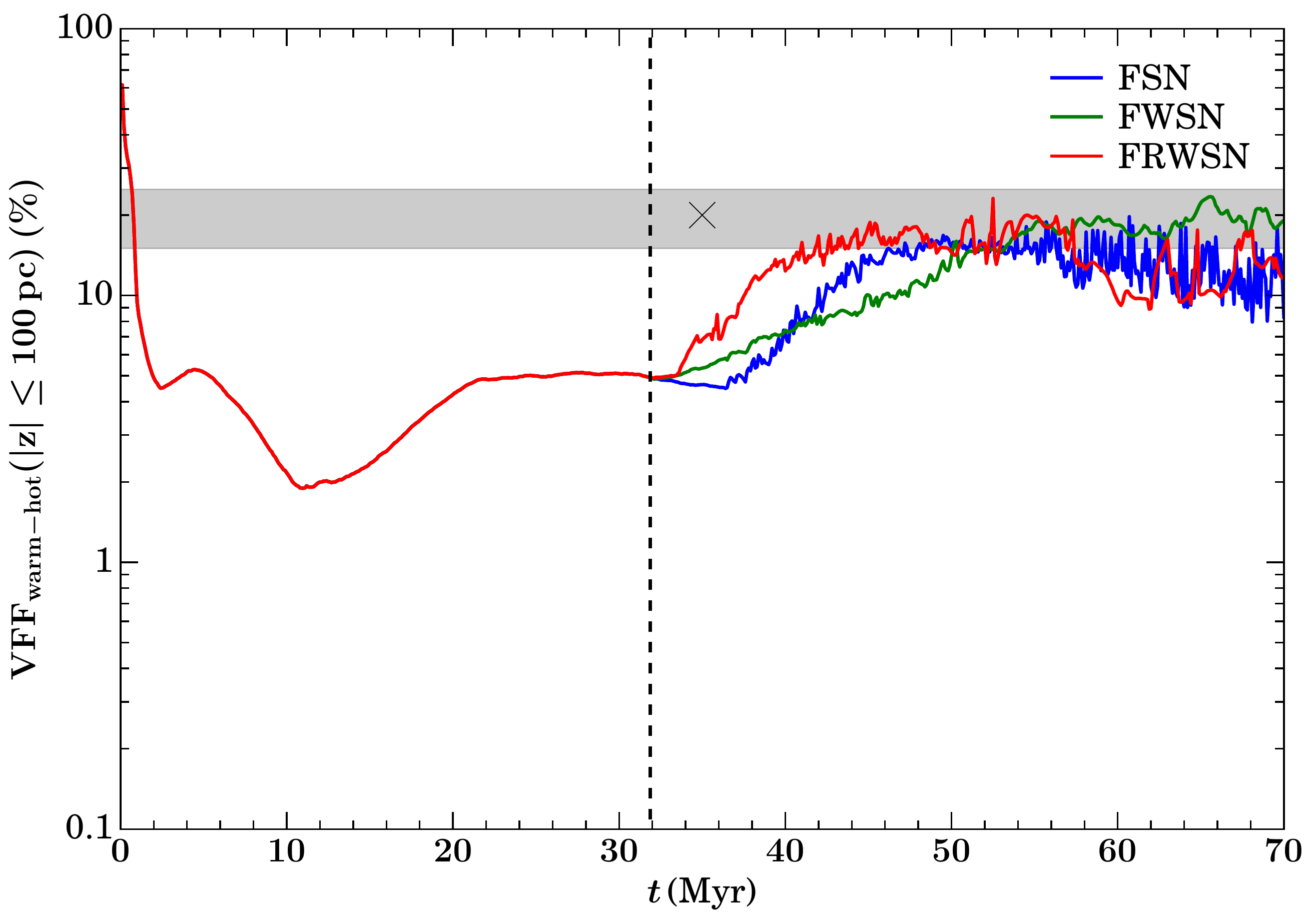}
\includegraphics[width=0.5\linewidth]{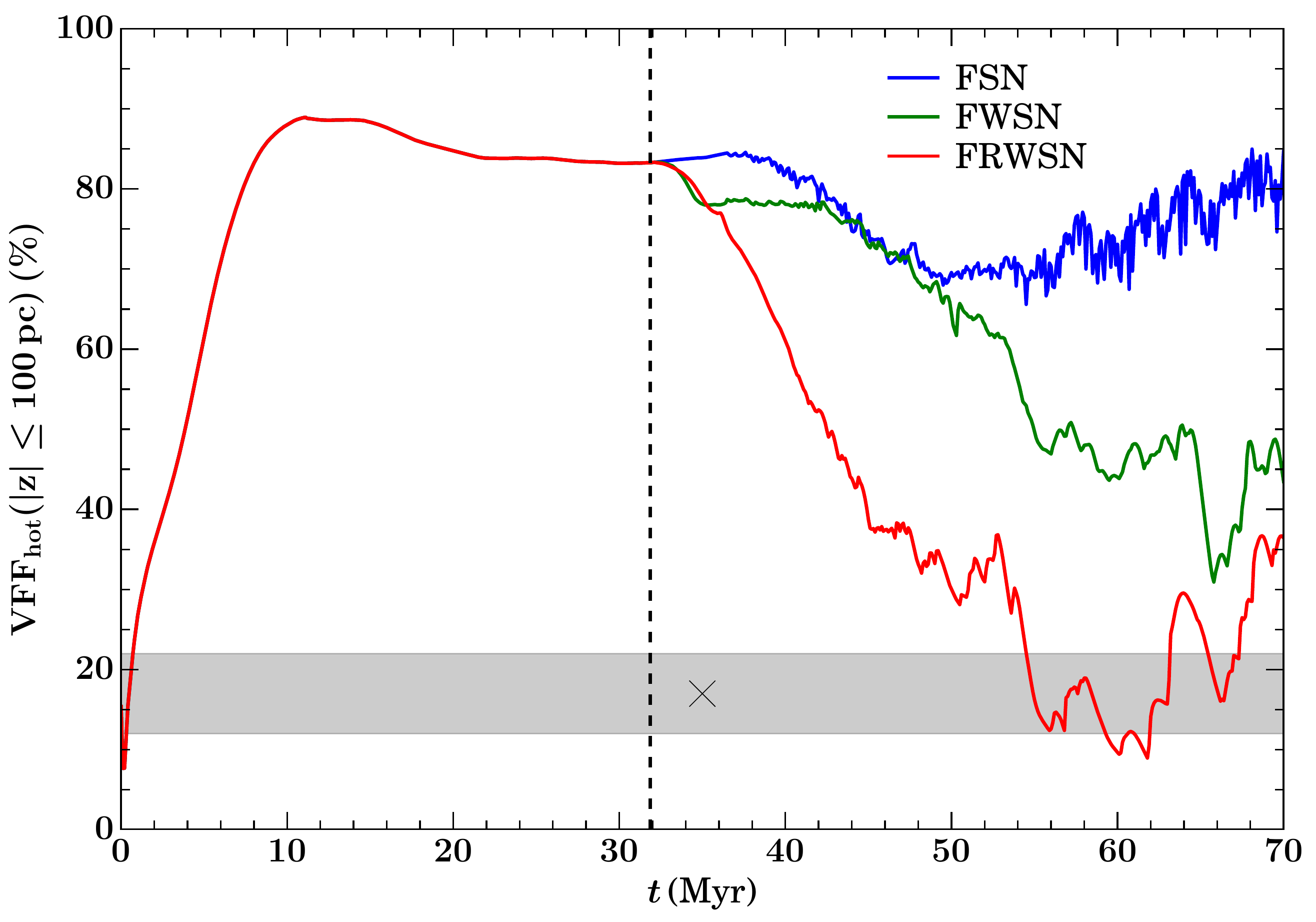}}
\caption{Volume filling fractions within the volume $\left| z\right| \leq 100\,$pc for the cold, warm, warm-hot and hot phase
as a function of time $t$ for all simulations (from top left to bottom right).
The vertical dashed line marks the onset of star formation. Crosses and grey bands are the data of \citet{kalker09}.}
\label{fig:vff}
\end{figure*}

Another interesting property of the ISM is the volume filling fraction of the different ISM phases.
Following \citet{gatto16}, we use these temperature cuts to define the phases:
\begin{enumerate}
\item molecular phase ($T \leq 30\,$K),
\item cold phase ($30 < T \leq 300\,$K),
\item warm phase ($300 < T \leq 8000\,$K),
\item warm-hot phase ($8000 \leq T \leq 3 \times 10^5\,$K),
\item hot phase ($T > 3 \times 10^5\,$K).
\end{enumerate}
Note that molecular here refers to the CO-bright molecular gas. CO-dark molecular gas has a much
broader temperature distribution \citep[see e.g.][]{glosmi16}, and in our simulations much of it is
located in our cold phase \citep[see][for more on this point]{walchetal15,peters16a}.
In our analysis, we ignore the molecular phase since it is not well resolved in our simulations.
The time evolution of the volume filling fractions of the cold, warm, warm-hot and hot phases
are shown in Figure~\ref{fig:vff}. We restrict the computation of the volume filling fraction
to the inner disc region with $\left| z\right| \leq 100\,$pc.

The volume filling fraction for the cold phase shows only little variation with time in all cases
and falls between $2\,$\% and $9\,$\%. The situation is similar for the warm-hot phase,
where the volume filling fraction lies between $8\,$\% and $11\,$\%. Here, there is no clear trend
between the simulations. This is understandable since the warm-hot phase is thermally unstable,
so we only see gas in a transient state.

The situation is different for the other two phases, which show a behaviour that is approximately opposite to each other.
Run FSN produces a volume filling fraction of the hot phase of $85\,$\% and of the warm phase of $10\,$\%.
Run FWSN, with a factor of ten smaller SFR, produces a volume filling fraction of the hot phase of only $50\,$\%,
while the volume filling fraction of warm phase increases to $30\,$\%. The hot phase volume filling fraction
in run FRWSN, with a factor of two smaller SFR, varies around $25\,$\%, and in the warm phase around $75\,$\%.
A comparison between the simulations is difficult because one has to separate the effects
of the different SFR from the impact of the various forms of stellar feedback. But the differences between
the simulations FWSN and FRWSN appear too large to be primarily a result of the smaller SFR in run FRWSN.
Instead, they are likely the result of the inclusion of radiative feedback in run FRWSN. Photoionisation
produces a lot more ionised gas than is present in run FWSN (compare Figure~\ref{fig:mf}), which then
radiatively cools and enters the warm phase. This is why the volume filling fraction in the warm phase is enhanced,
at the cost of the hot phase volume filling fraction.
\citet{gatto16} have shown that a volume filling fraction in excess of $50\,$\% in the hot phase
is necessary to drive galactic outflows \citep[cosmic rays modify this picture, see e.g.][]{2015ApJ...813L..27P,giricr,simps16}.
This is consistent with our finding that only run FSN launches an outflow
that leaves the inner $\pm 1\,$kpc of the box in the vertical direction during the simulation.

Our results can be compared with the observation-based models for the ISM phase volume filling fractions
presented in \citet{kalker09}. In the inner $100\,$pc, they find a volume filling fraction in the 
cold phase of $18\,$\%, in the warm phase of $42\,$\%, in the warm-hot phase of $20\,$\%, and in the
hot phase of $17\,$\% (their Figure~11). These values are shown as crosses in Figure~\ref{fig:vff}.
The cold phase is the only phase where run FRWSN does not reproduce the observed value, in the simulation it is too small by a factor
of three. The observed volume filling fraction of the warm-hot phase is matched by all simulations.
But for the warm and hot phases, the simulation with radiation is the only one that approaches the oberved
values. The other simulations underestimate the warm and overerstimate the hot phase.

\section{H$\alpha$ maps}
\label{sec:halpha}

The ionised gas in the ISM can be observed with the H$\alpha$ recombination radiation emitted
during the Balmer transition $n = 3 \to n = 2$. The H$\alpha$ radiation
is mostly emitted from gas with $T \approx 10^4\,$K, which is primarily photoionised gas. The gas that is shock-ionised
by winds and supernovae is typically too hot to be observed in H$\alpha$, but can become visible as the shocks cool down.
Because of the close connection between H$\alpha$ emission and \hii\ regions from massive stars, H$\alpha$ is an important SFR tracer.
Our simulations allow us to investigate systematic errors in the calibration of SFR measurements in H$\alpha$, and
in particular to quantify the contamination of the H$\alpha$ flux from shock emission.

To this end, we produce synthetic H$\alpha$ observations of our simulation box.
Two processes contribute to the emission of H$\alpha$ radiation. The first one is the recombination of ionised hydrogen
with a free electron. We describe the emissivity caused by radiative recombination following Eq.~(9) of \citet{dondra11} as
\begin{equation}
j_{\mathrm{H}\alpha\mathrm{, R}} = 2.82 \times 10^{-26}\, T_4^{-0.942-0.031 \mathrm{ln}(T_4)}\, n_\mathrm{e} n_\mathrm{H^+} \frac{\mathrm{\, erg}}{\mathrm{cm}^3\mathrm{\, s\, sr}} ,
\label{eq:jr}
\end{equation}
where $n_\mathrm{e}$ and $n_\mathrm{H^+}$ are the number densities of electrons and protons, respectively, and
$T_4 = T / 10^4\,$K with the gas temperature $T$. The second contribution comes from collisional excitation of neutral
hydrogen by free electrons. We follow \citet{kimetal13b} and set
\begin{equation}
j_{\mathrm{H}\alpha\mathrm{, C}} = \frac{1.30 \times 10^{-17}}{4 \pi} \frac{\Gamma_{13}(T)}{\sqrt{T}} \exp\left(\frac{-12.1\,\mathrm{eV}}{k_\mathrm{B}T}\right)
n_\mathrm{e} n_\mathrm{H} \frac{\mathrm{\, erg}}{\mathrm{cm}^3\mathrm{\, s\, sr}}
\label{eq:jc}
\end{equation}
according to their Eq.~(6), with Boltzmann's constant $k_\mathrm{B}$, the number density of neutral hydrogen $n_\mathrm{H}$ and
the Maxwellian-averaged effective collision strength
\begin{equation}
\Gamma_{13}(T) = 0.35 - 2.62 \times 10^{-7} T - 8.15 \times 10^{-11} T^2 + 6.19 \times 10^{-15} T^3 ,
\end{equation}
for $4,000\,$K\,$\leq T \leq 25,000\,$K and
\begin{equation}
\Gamma_{13}(T) = 0.276 + 4.99 \times 10^{-6} T - 8.85 \times 10^{-12} T^2 + 7.18 \times 10^{-18} T^3 ,
\end{equation}
for $25,000\,$K\,$< T \leq 500,000\,$K. The formulas for the collision strength are based on polynomial interpolation
through data by \citet{aggarwal83}.
In the subsequent discussion, we will look at these two contributions separately as well as at their
sum $j_{\mathrm{H}\alpha} = j_{\mathrm{H}\alpha\mathrm{, R}} + j_{\mathrm{H}\alpha\mathrm{, C}}$.
For simplicity, we assume $n_\mathrm{e} = n_\mathrm{H^+}$ during post-processing.
Deviations from this (due to e.g. the presence of helium) will change our values by less than $20$\%.

\begin{figure}
\centerline{\includegraphics[width=\linewidth]{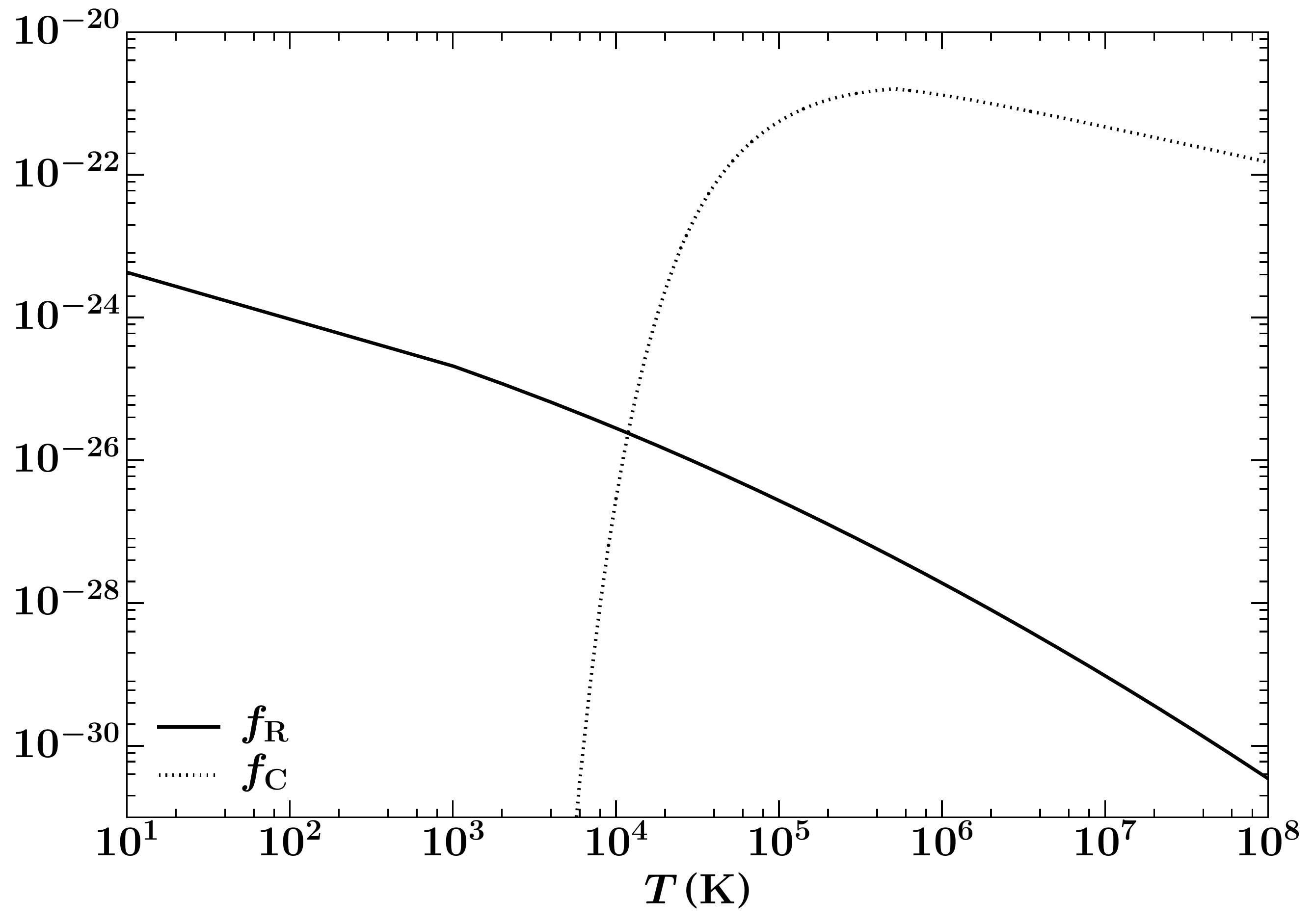}}
\caption{Density-independent prefactors $f_\mathrm{R}$ and $f_\mathrm{C}$
of radiative recombination emission and collisional excitation emission, respectively,
as a function of temperature $T$.}
\label{fig:functions}
\end{figure}

Both emissivities $j_{\mathrm{H}\alpha\mathrm{, R}}$ and $j_{\mathrm{H}\alpha\mathrm{, C}}$ have a similar functional form.
Radiative recombination emission is proportional to the product $n_\mathrm{e} n_\mathrm{H^+}$ with a temperature-dependent prefactor $f_\mathrm{R}(T)$,
whereas collisional excitation emission is proportional to $n_\mathrm{e} n_\mathrm{H}$ with prefactor $f_\mathrm{C}(T)$.
The temperature-dependence of these proportionality factors $f_\mathrm{R}(T)$ and $f_\mathrm{C}(T)$, which are
defined by the formulas~\eqref{eq:jr} and \eqref{eq:jc}, respectively, 
is shown in Figure~\ref{fig:functions}. 
Assuming that the products of number densities are of a similar magnitude, radiative
recombination dominates at temperatures $T \ll 10^4\,$K, whereas collisional excitation becomes dominant at $T \gg 10^4\,$K.
The fact that H$\alpha$ emission primarily originates from gas with $T \approx 10^4\,$K is because of the temperature-dependence
of the different ionisation stages of hydrogen. The abundance of $n_\mathrm{H}$ drops substantially above $T \approx 10^4\,$K,
reducing the magnitude of $j_{\mathrm{H}\alpha\mathrm{, C}}$. Likewise, the abundance of $n_\mathrm{H^+}$ decreases signifantly below
$T \approx 10^4\,$K, which diminishes $j_{\mathrm{H}\alpha\mathrm{, R}}$.

We integrate the emissivities along parallel rays through the simulation box, neglecting scattering and absorption by
intervening dust. For the projection along the $z$-axis, which mimics a patch of a face-on view of a Milky Way-type galaxy,
we compute the total H$\alpha$ luminosity by integrating over the entire image.
We will also separately consider the H$\alpha$ luminosity from radiative recombination emission
and from collisional excitation emission only.

\section{H$\alpha$ luminosity surface density}
\label{sec:lsd}

\begin{figure}
\includegraphics[width=\linewidth]{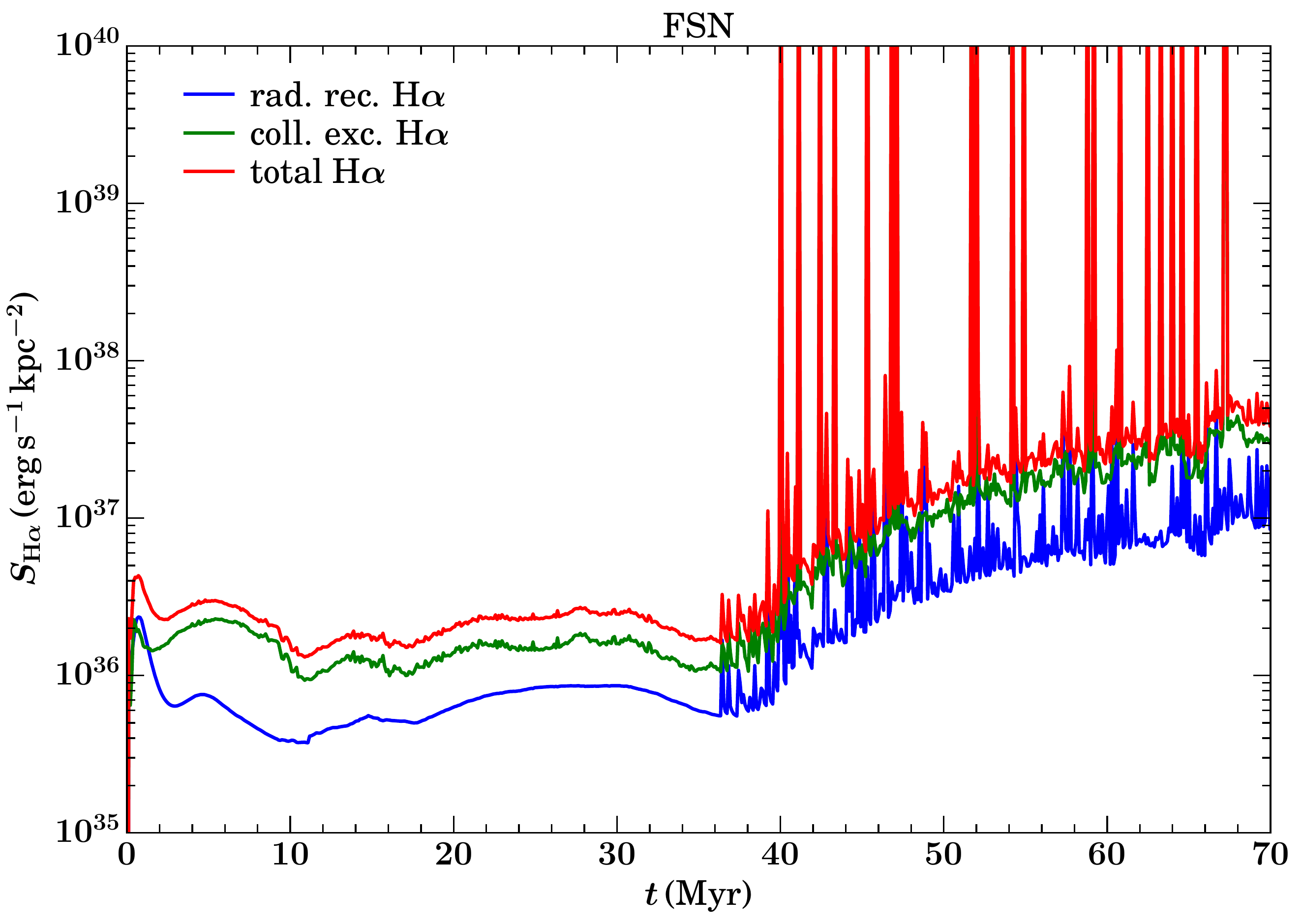}
\includegraphics[width=\linewidth]{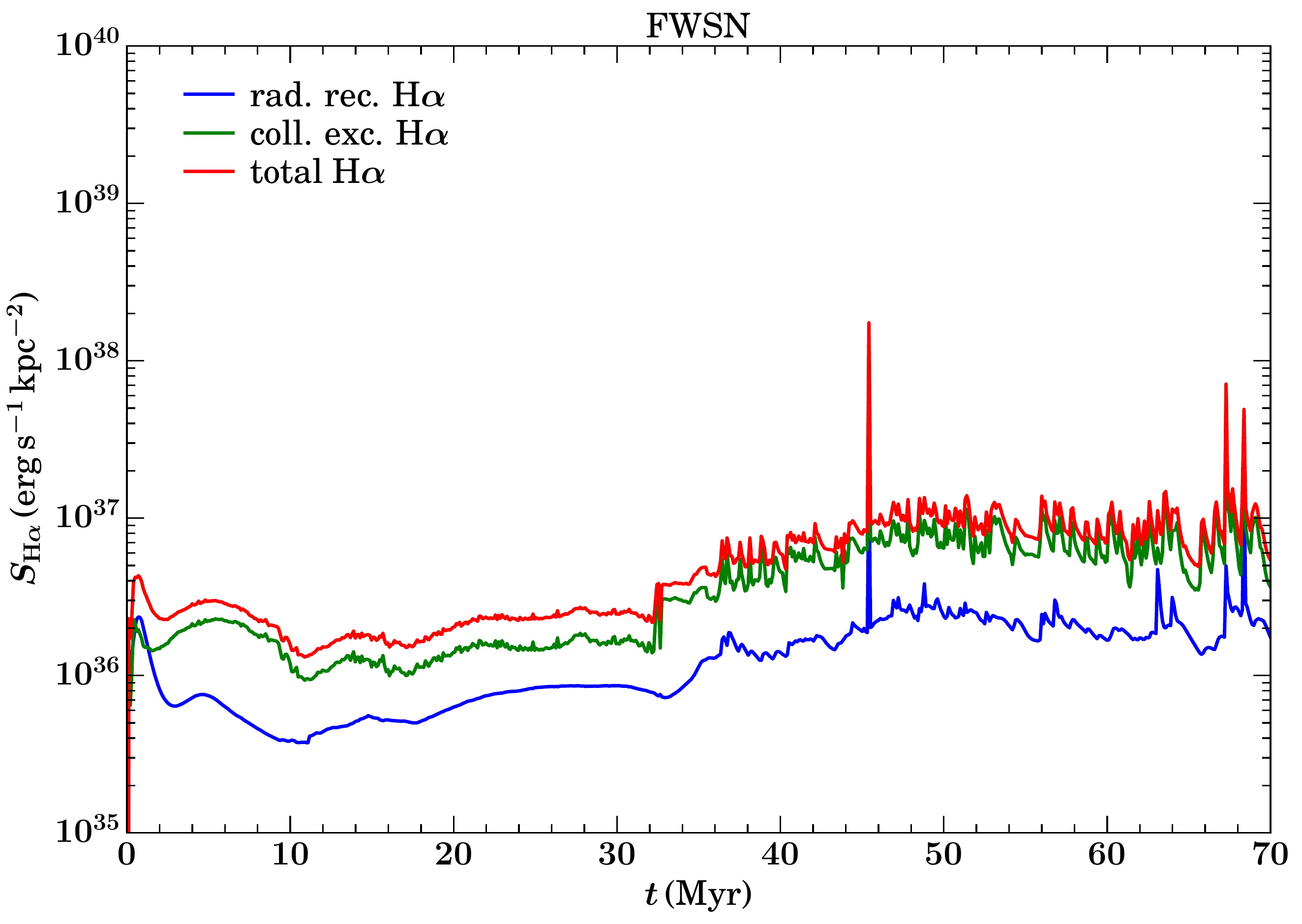}
\includegraphics[width=\linewidth]{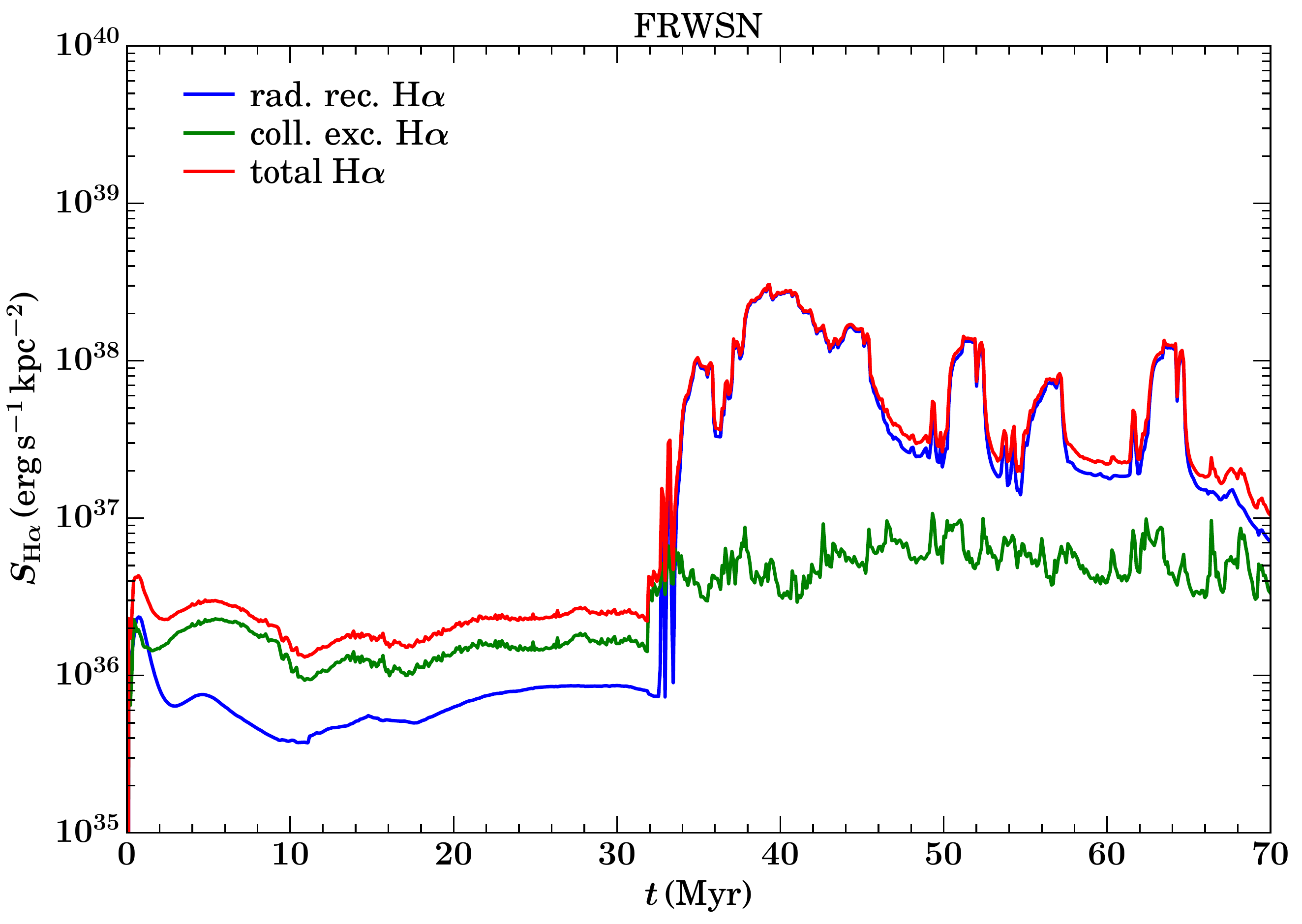}
\caption{H$\alpha$ luminosity surface density $S_{\mathrm{H}\alpha}$ for all simulations as a function of time $t$. The plots show the
contribution from radiative recombination emission, collisional excitation emission and the total emission.}
\label{fig:halsd}
\end{figure}

In Figure~\ref{fig:halsd} we show the H$\alpha$ luminosity surface density $S_{\mathrm{H}\alpha}$ for the face-on images
for the three simulations as a function of time $t$. This data is identical to the maps shown in the movies corresponding to
the Figures~\ref{fig:frwsn}, \ref{fig:fwsn} and \ref{fig:fsn}, respectively. We plot the contribution from radiative recombination
emission, collisional excitation emission, and the total emission.

In run FSN, $S_{\mathrm{H}\alpha}$ grows with time after the onset of the supernova explosions. Superimposed on
this average growth is a series of very pronounced spikes of the H$\alpha$ emission that boost the H$\alpha$ luminosity
by many orders of magnitude. These spikes are produced
by the injection of thermal energy into the dense gas inside the sink particle radius during a supernova explosion, which produces a large
amount of collisional ionisation. These injections are visible as bright spots around the sink particles in the H$\alpha$ panels
of the corresponding movie. Collisional excitation emission dominates over radiative recombination emission by a factor of $2$--$3$ on average.

In run FWSN, stellar winds evacuate the sink particle volume before the first supernova explosions commence. Thus,
the thermal energy injection during supernova explosions create much less collisional ionisation, and as
a consequence the spikes in $S_{\mathrm{H}\alpha}$ disappear. We already know that in this simulation, stellar
feedback does not produce much ionised gas (compare Figure~\ref{fig:mf}), and therefore $S_{\mathrm{H}\alpha}$
only increases by a factor of a few after the onset of star formation, but remains roughly constant throughout the simulation. Again, collisional
excitation emission dominates over radiative recombination emission by a small factor.

For run FRWSN, the situation is very different. We have already seen in Figure~\ref{fig:mf} that the radiative feedback
boosts the amount of ionised gas by the formation of \hii\ regions. Figure~\ref{fig:halsd} demonstrates that, as
a consequence, $S_{\mathrm{H}\alpha}$ increases by one order of magnitude compared to run FWSN. Interestingly,
the flux coming from collisional excitation emission remains approximately at the same level as in run FWSN, whereas the
rise in the H$\alpha$ flux is entirely due to a much enhanced radiative recombination emission. This is the radiation
coming from the \hii\ regions. As soon as the first \hii\ regions form, radiative recombination emission always dominates over
collisional excitation emission, but the difference can vary from a factor of only $2$ up to more than an order of magnitude.
If we identify collisional excitation emission with shocks and radiative recombination emission with \hii\ regions, then
shocks can contribute at most $30\,$\% to the total H$\alpha$ flux, but typically less than $10\,$\%. The total H$\alpha$
flux varies in the same way as the mass fraction of ionised gas because of the oscillations in the radiative energy
output (compare Figure~\ref{fig:erad}).

\section{Star formation rate calibration}
\label{sec:cal}
Measurements of the H$\alpha$ flux are routinely used as SFR tracers. To this end,
the H$\alpha$ luminosity surface density $S_{\mathrm{H}\alpha}$ is converted into an SFR
surface density $\Sigma_{\mathrm{SFR}}$. The typical conversion factors between
$S_{\mathrm{H}\alpha}$ and $\Sigma_{\mathrm{SFR}}$ used in the literature only differ by a factor of $2$.
We consider the following calibrations:
\begin{itemize}
\item $\Sigma_{\mathrm{SFR}} = 8.9 \times 10^{-42}\, S_{\mathrm{H}\alpha}$ \citep{kenni83},
\item 
$\Sigma_{\mathrm{SFR}} = 7.9 \times 10^{-42}\, S_{\mathrm{H}\alpha}$ \citep{kenn98},
\item $\Sigma_{\mathrm{SFR}} = 5.3 \times 10^{-42}\, S_{\mathrm{H}\alpha}$ \citep{calzetti07},
\item $\Sigma_{\mathrm{SFR}} = 5.45 \times 10^{-42}\, S_{\mathrm{H}\alpha}$ \citep{calzetti10}.
\end{itemize}
In these formulae, $S_{\mathrm{H}\alpha}$ is assumed to be given in units erg$\,$s$^{-1}\,$kpc$^{-2}$,
so that $\Sigma_{\mathrm{SFR}}$ has units M$_\odot\,$yr$^{-1}\,$kpc$^{-2}$.

SFR calibrations including H$\alpha$ \citep{kenn98}
connect the observed emssion of galaxies to the reprocessed light from
a population of young massive stars by the ISM. These calibrations rely on a
variety of assumptions, including the IMF, and number of ionising photons
absorbed by dust, which do not produce H$\alpha$. The rates used here are derived 
using extinction-corrected H$\alpha$ fluxes, with assumptions about the geometry of
sources and dust \citep{calzetti07}. A future improvement of our analysis will include dust absorption
of ionising radiation, which would lower the calculated emission per unit of SFR.
The column densities of this simulation are similar to the galaxies in \citet{boq16}.
There they found the amount of Lyman continuum photons absorbed by dust to be only $10\,$\%.

\begin{figure}
\includegraphics[width=\linewidth]{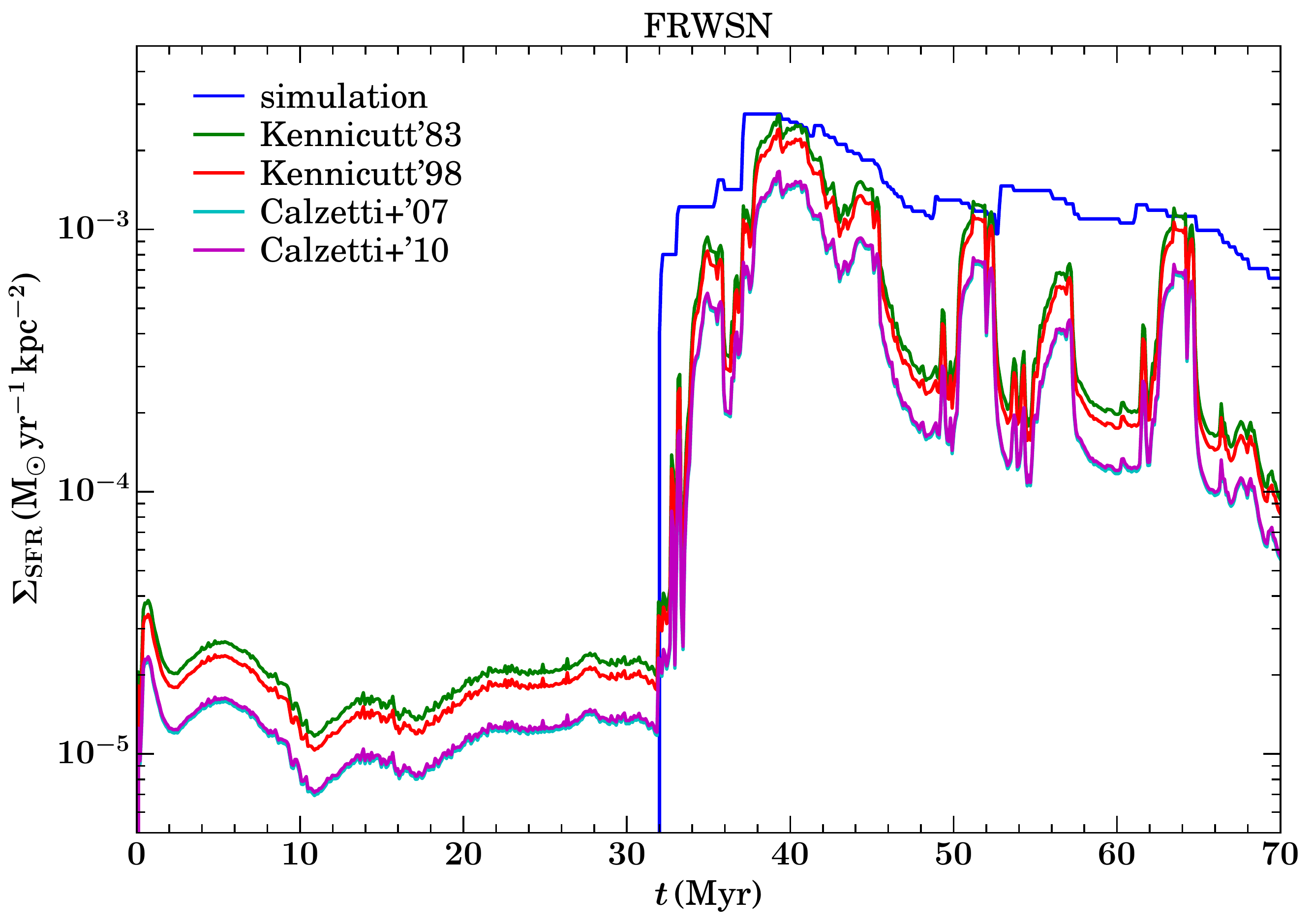}
\caption{SFR surface density $\Sigma_{\mathrm{SFR}}$ measured in the simulation and derived via
various H$\alpha$ SFR calibrations as a function of time $t$ for run FRWSN.}
\label{fig:ks}
\end{figure}

Figure~\ref{fig:ks} shows the SFR measured in H$\alpha$ using these calibrations for run FRWSN
together with the true SFR, which we have already computed in Section~\ref{sec:stcl} (compare Figure~\ref{fig:sfrOB}).
As discussed previously, the true SFR was derived by distributing the mass of a newly born cluster
over the associated massive star's lifespan, thereby taking into account the period over which it emits ionising radiation.
This SFR contrasts with the naive theoretical SFR shown in Figure~\ref{fig:sfr}, which simply bins star formation events in
$1\,$Myr intervals and does not take the finite stellar lifetime that affects observational SFR measurements into account.
We therefore expect a good agreement between our true SFR and the SFR measured in H$\alpha$.

Indeed, after the onset of star formation, the observed and the true SFR follow each other closely. However,
after $45\,$Myr the H$\alpha$ flux drops by an order of magnitude, whereas the true SFR remains roughly constant.
In the following, we see a series of oscillations in the observed SFR that has no correspondence in the true SFR.
We have already found that the origin of these oscillations is the death of very massive stars
with $M \geq 30\,$M$_\odot$ that dominate the stellar luminosities, but have short lifetimes (compare Figure~\ref{fig:eradbin}).
The true SFR does not oscillate because most stars in the clusters are less massive and therefore long-lived
(compare Figure~\ref{fig:hist}).
This plot therefore demonstrates that H$\alpha$ measurements of the SFR are only accurate when very massive stars
are present. Less massive stars do not produce enough ionising flux to produce an H$\alpha$ emission that matches their
SFR. In this case, the H$\alpha$ measurement can underestimate the SFR by an order of magnitude, independent of the calibration
used. This is an important
systematic error for H$\alpha$ measurements of the SFR \citep[see also][]{dasilva14,hony15}.

\section{Conclusions}
\label{sec:con}

We present simulations of the multi-phase ISM that simultaneously include stellar feedback in the form of supernovae, stellar winds and radiation.
These chemo-radiation-hydrodynamical simulations are part of the SILCC project and extend previous simulations
of self-regulated star formation by winds and supernovae \citep{gatto16}
by radiative feedback in the form of photoelectric heating, photodissociation and photoionisation from star clusters
using the \texttt{Fervent} code \citep{baczynski15}.

We find that photoionisation feedback contributes to the regulation of star formation by an increase in thermal pressure
that the accretion flow around the star cluster must overcome to continue growing. Therefore, a simulation with feedback
by radiation, winds and supernovae has on average lower-mass clusters that accrete for a shorter period of time compared to a simulation
with only wind and supernova feedback. As a consequence, the SFR is reduced by a factor of $2$ in this case.
For the same reason, supernovae explode in an environment with a lower mean density in the presence
of radiation.

We find that, for this simulation, photoionisation heating is the dominant energy source in the ISM and that it exceeds the energy input from supernovae
by one and from winds by two orders of magnitude. All photochemical processes can individually impart more energy into the
ISM than supernovae, provided that this radiation is absorbed by the material. The cluster luminosities are highly
variable with time because they are dominated by very massive stars ($M \geq 30\,$M$_\odot$) with lifetimes of only a few Myr.

The presence of radiative feedback significantly affects the mass fractions of the different chemical states of hydrogen.
The mass fractions of atomic and ionised hydrogen increase whereas the molecular hydrogen mass fraction drops.
Photoionisation by star clusters is the dominant source of ionised gas in the ISM. This ionised gas then cools radiatively
and produces a much enhanced volume filling fraction of the warm phase and a substantial reduction of the hot phase volume
filling fraction compared to simulations without radiation. This is essential to match the observed volume filling fractions
of the warm and hot phases.
The simulation with radiation naturally exhibits a depletion time that
is in agreement with observations, while the other simulations fail to do so.

The time variability of the cluster luminosities has important consequences for SFR measurements in H$\alpha$.
We find that the SFR observed in H$\alpha$ only matches the true SFR when very massive stars are present in the clusters.
Less massive stars do not produce enough ionising radiation to create an H$\alpha$ flux that matches their SFR.
The H$\alpha$ measurement then underestimates the SFR by up to an order of magnitude, and this result is independent
of the calibration used. Shock emission typically contributes less than $10\,$\% to the total H$\alpha$ flux,
but can go up to $30\,$\%.

\section*{Acknowledgements}

All simulations have been performed on the Odin and Hydra clusters hosted by the Max Planck Computing \& Data Facility (http://www.mpcdf.mpg.de/).
TP, TN, SW, SCOG, PG and RSK acknowledge the {\em Deutsche Forschungsgemeinschaft (DFG)} for funding through the SPP 1573
``The Physics of the Interstellar Medium''.
TN acknowledges support by the DFG cluster of excellence ``Origin and structure of the Universe''. 
SW acknowledges funding by the Bonn-Cologne-Graduate School,
by SFB 956 ``The conditions and impact of star formation'', and from the European Research Council under the European Community's Framework
Programme FP8 via the ERC Starting Grant RADFEEDBACK (project number 679852).
SCOG and RSK acknowledge support from the DFG via SFB 881 ``The Milky Way System'' (sub-projects B1, B2 and B8).
RSK acknowledges support from the European Research Council under the European Community's Seventh Framework Programme
(FP7/2007-2013) via the ERC Advanced Grant STARLIGHT (project number 339177).
RW acknowledges support by project 15-06012S of the Czech
Science Foundation and the institutional project
RVO:~67985815.
The software used in this work was developed in part by the DOE NNSA ASC- and DOE Office of Science ASCR-supported Flash Center for
Computational Science at the University of Chicago.
The data analysis was partially carried out with the \texttt{yt} software \citep{turketal11}.

\end{document}